\documentclass[longauth]{aa} 

\usepackage{longtable,lscape}
\usepackage{amsmath}
\usepackage{natbib}
\bibpunct{(}{)}{;}{a}{}{,}
\usepackage{graphicx}
\usepackage{multirow}

\usepackage{graphicx}
\usepackage{txfonts}
\usepackage{natbib}
\usepackage[colorlinks=true,citecolor=blue]{hyperref}
\usepackage{color}
\usepackage{xspace}
\usepackage{dcolumn}
\usepackage{longtable}
\usepackage{lscape}
\usepackage{soul}        

\usepackage{rotating}
\usepackage{afterpage}

\newcommand{\MJup}{M$_{\mathrm{Jup}}$\xspace}

\newcommand{\as}{\hbox{$^{\prime\prime}$}\xspace}

\usepackage{txfonts}
\begin{document}
\title{First light of the VLT planet finder SPHERE}
   \subtitle{IV. Physical and chemical properties of the planets around HR8799\thanks{Based on observations collected at the European Southern Observatory, Chile, during the commissioning of the SPHERE instrument}}

   \author{M. Bonnefoy\inst{1}, A. Zurlo\inst{2,3}, J.L. Baudino\inst{4},  P. Lucas\inst{5}, D. Mesa\inst{6},  A.-L. Maire\inst{6, 7}, A. Vigan\inst{8, 9}, R. Galicher\inst{4}, D. Homeier\inst{10,11}, F. Marocco \inst{5}, R. Gratton\inst{6}, G. Chauvin\inst{1}, F. Allard\inst{11}, S. Desidera\inst{6}, M. Kasper\inst{12,1}, C. Moutou\inst{8}, A.-M. Lagrange\inst{1},  A. Baruffolo\inst{6}, J. Baudrand\inst{4}, J.-L. Beuzit\inst{1}, A. Boccaletti\inst{4}, F. Cantalloube\inst{1,13}, M. Carbillet\inst{14}, J. Charton\inst{1}, R.U. Claudi\inst{6}, A. Costille\inst{8}, K. Dohlen\inst{13}, C. Dominik\inst{15}, D. Fantinel\inst{6}, P. Feautrier\inst{1}, M. Feldt\inst{7}, T. Fusco\inst{13,6}, P. Gigan\inst{4}, J. H. Girard\inst{9,1}, L. Gluck\inst{1}, C. Gry\inst{8}, T. Henning\inst{7}, M. Janson\inst{16,7}, M. Langlois\inst{11, 8}, F. Madec\inst{8}, Y. Magnard\inst{1}, D. Maurel\inst{1}, D. Mawet\inst{9,17}, M. R. Meyer\inst{18}, J. Milli\inst{9,1}, O. Moeller-Nilsson\inst{7}, D. Mouillet\inst{1}, A. Pavlov\inst{7}, D. Perret\inst{4}, P. Pujet\inst{1}, S. P. Quanz\inst{18}, S. Rochat\inst{1}, G. Rousset\inst{4}, A. Roux\inst{1}, B. Salasnich\inst{6}, G. Salter\inst{8}, J.-F. Sauvage\inst{6, 13}, H.M. Schmid\inst{18}, A. Sevin\inst{4}, C. Soenke\inst{12}, E. Stadler\inst{1}, M. Turatto\inst{6}, S. Udry\inst{19}, F. Vakili\inst{14}, Z. Wahhaj\inst{9,8}, F. Wildi\inst{19}}
 \institute{\inst{1} Univ. Grenoble Alpes, IPAG, F-38000 Grenoble, France. CNRS, IPAG, F-38000 Grenoble, France \\
 \inst{2} N\'{u}cleo de Astronom\'{i}a, Facultad de Ingenier\'{i}a, Universidad Diego Portales, Av. Ejercito 441, Santiago, Chile \\
 \inst{3} Millenium Nucleus ``Protoplanetary Disk", Departamento de Astronom\'{i}a, Universidad de Chile, Casilla 36-D, Santiago, Chile \\
 \inst{4} LESIA, CNRS, UMR 8109, Observatoire de Paris, Univ. Paris Diderot, UPMC, 5 place Jules Janssen, 92190 Meudon, France \\
 \inst{5} Centre for Astrophysics Research, Science and Technology Research Institute, University of Hertfordshire, Hatfield AL10 9AB, United Kingdom\\
 \inst{6} INAF-Osservatorio Astronomico di Padova, Vicolo dell'Osservatorio 5, 35122, Padova, Italy \\               
 \inst{7} Max-Planck-Institut f\"{u}r Astronomie, K\"{o}nigstuhl 17, 69117 Heidelberg, Germany\\
 \inst{8} Aix Marseille Universit\'{e}, CNRS, LAM (Laboratoire d'Astrophysique de Marseille) UMR 7326, 13388, Marseille, France\\
  \inst{9} European Southern Observatory, Alonso de Cordova 3107, Vitacura, Santiago, Chile\\
  \inst{10} Zentrum f\"{u}r Astronomie der Universit\"{a}t Heidelberg, Landessternwarte K\"{o}nigstuhl 12, 69117 Heidelberg, Germany \\
  \inst{11} CRAL, UMR 5574, CNRS, Universit\'{e} de Lyon, Ecole Normale Sup\'{e}rieure de Lyon, 46 All\'{e}e d’Italie, F-69364 Lyon Cedex 07, France\\ 
  \inst{12} European Southern Observatory, Karl-Schwarzschild-Strasse 2, D-85748 Garching, Germany \\ 
  \inst{13} ONERA - The French Aerospace Lab BP72 - 29 avenue de la Division Leclerc FR-92322 Chatillon Cedex\\
  \inst{14} Laboratoire Lagrange, UMR7293, Universit\'e de Nice Sophia-Antipolis, CNRS, Observatoire de la Cote d'Azur, Bd. de l'Observatoire, 06304 Nice, France\\
  \inst{15} Anton Pannekoek Astronomical Institute, University of Amsterdam, PO Box 94249, 1090 GE Amsterdam, The Netherlands\\
   \inst{16} Stockholm University, AlbaNova University Center, Stockholm, Sweden \\ 
  \inst{17} Department of Astronomy, California Institute of Technology, 1200 E. California Blvd, MC 249-17, Pasadena, CA 91125 USA\\
  \inst{18} Institute for Astronomy, ETH Zurich, Wolfgang-Pauli-Strasse 27, 8093 Zurich, Switzerland \\
  \inst{19} Observatoire de Gen\'{e}ve, University of Geneva, 51 Chemin des Maillettes, 1290, Versoix, Switzerland}

   \date{Received 06 July 2015 / Accepted 15 September 2015}


  \abstract
   {The system of four planets discovered around the intermediate-mass star HR8799 offers a unique opportunity to test planet formation theories at large orbital radii and to probe the physics and chemistry at play in the atmospheres of self-luminous young ($\sim$30 Myr) planets. We recently obtained new photometry of the four planets and low-resolution (R$\sim$30) spectra of HR8799 d and e with the SPHERE instrument (paper III).}
   {In this paper (paper IV), we aim to use these spectra and available photometry to determine how they compare to known objects, what the planet physical properties are, and how their atmospheres work.}
   {We  compare the available spectra, photometry, and spectral-energy distribution (SED) of the planets to field dwarfs and young companions. In addition, we use the extinction from corundum, silicate (enstatite and forsterite), or iron grains likely  to form in the atmosphere of the planets to try to better understand empirically the peculiarity of their spectrophotometric properties. To conclude, we use three sets of atmospheric models (BT-SETTL14, Cloud-AE60, Exo-REM) to determine  which ingredients are critically needed in the models to represent the SED of the objects, and to constrain their atmospheric parameters ($T\mathrm{_{eff}}$, log $g$, M/H).}
   {We find that HR8799d and e properties are well reproduced by those of L6-L8 dusty dwarfs discovered in the field, among which some are candidate members of young nearby associations. No known object reproduces well the properties of planets b and c. Nevertheless, we find that the spectra and WISE photometry of peculiar and/or young early-T dwarfs reddened by submicron grains made of corundum, iron, enstatite, or forsterite successfully reproduce the SED of these planets. Our analysis confirms that  only the Exo-REM  models with thick clouds  fit (within 2$\sigma$)  the whole set of spectrophotometric datapoints available for HR8799 d and e for $T\mathrm{_{eff}=1200}$ K, log $g$ in the range 3.0-4.5, and M/H=+0.5. The models still fail to reproduce the SED of HR8799c and b. The determination of the metallicity, log $g$, and cloud thickness are degenerate.}
{Our empirical analysis and atmospheric modelling show that an enhanced content in dust and decreased CIA of $H_{2}$ is certainly responsible for the deviation of the properties of the planet with respect to field dwarfs. The analysis suggests in addition that HR8799c and b have later spectral types than the two other planets, and therefore could both have lower masses.}


\titlerunning{Physical and chemical properties of the planets around HR8799}
\authorrunning{Bonnefoy et al.}
\maketitle
%
\section{Introduction}
Direct imaging (DI) of nearby stars represents the only viable method to complete our view of the planetary system architectures, related formation mechanisms, and dynamical evolution at large ($>$ 5 AU) separations.  Nowadays, more than 40 planetary mass companions have been detected using DI\footnote{http://www.exoplanet.eu/}. Most of them are found orbiting at large separations ($\gg$ 50 AU) around young ($<$ 300 Myr)  K- and M-type stars. The large projected separations and high mass ratio between the system components, suggest that  alternative formation mechanisms to the classical Core-accretion paradigm \citep{1996Icar..124...62P} are responsible for this population of objects. 

Nonetheless, the favorable contrast of these companions with their host stars and the angular separation of the systems on sky make them amenable to follow-up  multi-band photometry and near-infrared (1-2.5 $\mu$m) spectroscopy. The companions are found to share the characteristics  of  young free-floating brown dwarfs (red near-infrared colors, triangular H-band shape, reduced alkali lines) discovered in  clusters, star forming regions, and young nearby associations \citep{2001MNRAS.326..695L, 2006ApJ...639.1120K, 2008MNRAS.383.1385L, 2010ApJ...715L.165R, 2012A&A...539A.151A, 2013ApJ...772...79A, 2013A&A...549A.123A, 2014ApJ...785L..14G, 2014ApJ...792L..17G, 2014A&A...562A.127B, 2014MNRAS.442.1586D, 2014A&A...572A..67Z}. Among them,  2MASSW J1207334-393254 b \citep[hereafter 2M1207b;][]{2004A&A...425L..29C}, 2MASS J01225093-2439505 b \citep{2013ApJ...774...55B}, and VHS J125601.92-125723.9 b \citep{2015ApJ...804...96G} appear to lie at the transition between the cloudy L-type and cloudless T-type dwarfs. Because they orbit stars with known distance, age, and (sometimes) metallicity, these rare young L/T-type companions  offer the  opportunity to study the key processes driving the disappearance of dust grains in the atmosphere of planets and brown dwarfs \citep[e.g.][]{2009ApJ...702..154S}.  Both 2MASSW J1207334-393254 b and  VHS J125601.92-125723.9 b are underluminous with respect to field dwarfs. Modelling with atmospheric models of the flux and spectral shape of these objets suggests that their atmospheres have a high content in atmospheric dust (thick clouds) and might be affected by non-equilibrium chemistry \citep{2011ApJ...735L..39B, 2011ApJ...732..107S,  2014ApJ...792...17S}.  The red spectra, colors and luminosity problem of these extreme objets are also observed for a growing population of isolated dusty L dwarfs identified in the field \citep[e.g.][]{2013ApJ...777L..20L, 2014AJ....147...34S, 2014MNRAS.439..372M, 2015ApJ...799..203G}.

The breaktrougth detections of four young, warm ($T\mathrm{_{eff}\sim1000}$K), and self-luminous planets around the  A-type star HR8799 \citep{2008Sci...322.1348M, 2010Natur.468.1080M} has opened a new window for the  constraint of planet formation processes and physics of gas giants soon after their
formation.  HR8799 (HD 218396; V432 Pegasi) is a $1.5 M_{\odot}$ star \citep{1999AJ....118.2993G, 2012ApJ...761...57B} located at a  distance of $39.4\pm1.0$ pc \citep{2007A&A...474..653V}. At the estimated age of the system \citep[$\sim$30~Myr,][]{2010Natur.468.1080M, 2012ApJ...761...57B, 2013ApJ...762...88M}, the luminosity of the four planets  yields estimated masses of $\sim$7, 7, 7, and 5 $M_{Jup}$ using "hot-start"  cooling tracks \citep{2003A&A...402..701B}. Their mass ratio with their host star  and short projected separations ($\sim$15--70 AU) distinguish them from the population of previously known planetary mass companions. The existence of a warm inner belt of debris and cold Kuiper belt bracketing the orbits of the planets \citep{2009ApJ...705..314S, 2011ApJ...740...38H,2014ApJ...780...97M,2015arXiv150202315C} indicates that the companions likely formed into a circumstellar disk. 

Beyond our Solar System, HR8799  provides the only sample of companions most likely originating from a "planet-like" formation process \citep{1996Icar..124...62P, 1997LPI....28..137B, 2012A&A...544A..32L}  that allow for comparative characterization. So far, studies of the atmospheres of HR\,8799\,bcde were conducted using mostly photometric measurements in the infrared (1-5 $\mu$m) \citep{2008Sci...322.1348M,2010Natur.468.1080M,2009ApJ...694L.148L,2010ApJ...716..417H,2010ApJ...710L..35J,
2011ApJ...729..128C,galicher11,2012ApJ...753...14S,2013A&A...549A..52E,2014ApJ...792...17S,
2014ApJ...795..133C}.  The measurements indicate that the four planets add to the restrained list of young companions at the L/T transition. They are also underluminous with respect to field dwarf counterparts \citep{2008Sci...322.1348M, 2010Natur.468.1080M}.

Similar to the case of 2M1207b, the planets spectrophotometry are better reproduced by models with thick clouds \citep[e.g.][]{
2011ApJ...733...65B, 2011ApJ...737...34M, 2012ApJ...754..135M}. Models with local variation of the cloud thickness  (“patchy" clouds) have been proposed to improve the fit of the planet spectral energy distributions   \citep[SEDs;][]{2011ApJ...729..128C, 2012ApJ...753...14S, 2014ApJ...795..133C}. At longer wavelengths, the planets show evidence for disequilibrium carbon chemistry, exhibiting weak  CH$_{4}$ absorption at 3.3 $\mu$m and strong CO absorption at 5 $\mu$m \citep{2010ApJ...716..417H, 2010ApJ...710L..35J, galicher11, 2012ApJ...753...14S, 2014ApJ...792...17S, 2014ApJ...795..133C}.

Near-infrared integral field spectroscopy (1-2.5 $\mu$m)  with the KECK/OSIRIS and P1640 instruments \citep{2010ApJ...723..850B, 2011ApJ...733...65B, 2015ApJ...804...61B, 2013Sci...339.1398K, 2013ApJ...768...24O}  provided a direct evidence for CO and H$_{2}$O absorption. They revealed the lack of strong CH$_{4}$ absorption, altough such absorption appears in the spectra of old field dwarfs at similar effective temperatures. This confirmed the efficient vertical mixing and disequilibrium CO/CH$_{4}$ chemistry in the atmosphere of the planets. These spectra along with the existing photometry were used to investigate whether traces of (differential) heavy element enrichment at formation could be retrieved in the atmosphere of HR8799b and c \citep{2013ApJ...778...97L, 2013Sci...339.1398K, 2015ApJ...804...61B} and help in discriminating between different formation pathways \citep[e.g.][]{2010Icar..207..503H, 2011ApJ...743L..16O}. 

Additional spectra are still needed in order to deepen our understanding of the physics and chemistry of these objects. The new generation of high-contrast imaging instruments such as SPHERE, GPI, and ScEXAO  \citep{2008SPIE.7014E..18B, 2014PNAS..11112661M,2009SPIE.7440E..0OM} can now provide in a single shot  the emission spectra of planets in the near-infrared (1-2.5 $\mu$m) at small angular separation without being limited by the contaminating stellar halo.  

\cite{2014ApJ...794L..15I} already presented K-band (1.9-2.4 $\mu$m) GPI low-resolution (R$\sim$60 to 80) spectra for planets c and d. In this paper we  take advantage of the new low-resolution (R$\sim$30) 0.98-1.64 $\mu$m spectra of HR8799d and e, and photometry of the four planets, obtained by SPHERE (Zurlo et al. submitted; hereafter Paper III), and of previously published photometry of the planets, to reinvestigate the properties of the planets. 

The outline of the paper is the following: in Section ~\ref{Sec:Data}, we give a summary of the photometric points available from literature and from the new SPHERE measurements used in this study. We conduct in Section ~\ref{subsec:empirical} an analysis of the spectral properties of the planets based on the comparison to known objects. We extend this analysis in Section \ref{subsec:model} with the use of atmospheric models. Our results are compared to previous studies and discussed in Section \ref{sec:discu}. We summarize our findings in Section~\ref{sec:conc}.

 \section{Data sample}
 \label{Sec:Data}
SPHERE provided  H2 (1.593 $\mu$m), H3 (1.667 $\mu$m), K1 (2.110  $\mu$m), and K2 (2.251 $\mu$m)  narrow-band  and J broad-band (1.245 $\mu$m) photometry of the four planets,  in addition to the 0.96-1.64 $\mu$m low-resolution spectra  of planets d and e. The K1-K2 photometry overlaps with the GPI spectra of HR8799c and d \citep{2014ApJ...794L..15I}, and the OSIRIS spectrum of HR8799b \citep{2011ApJ...733...65B}. For the latter, we considered the re-extracted spectrum of the planet presented in \cite{2015ApJ...804...61B}. The Keck He I B,  H$_{2}$($\nu$=1-0), Br$_{\gamma}$, and H$_{2}$ ($\nu$=2-1) photometry of HR8799b provided in \cite{2011ApJ...733...65B} is also redundant with the OSIRIS spectrum.  The J-band photometry of HR8799d and e  is  in agreement  with the SPHERE spectra. 

We attempted to add to the analysis the P1640 IFS spectra of HR8799b and c originally obtained by \cite{2013ApJ...768...24O} and re-extracted by \cite{2015ApJ...803...31P}. The P1640 spectrum of HR8799b appears to be dominated by noise below 1.4 $\mu$m and is super-seeded by the one obtained by \cite{2011ApJ...733...65B} at longer wavelengths. We did not use it for that reason. Conversely, the spectrum of HR8799c was flux-calibrated using the SPHERE J-band photometry and used in our analysis. 

We  completed these data shortward of 2.5 $\mu$m with Subaru $z$ band photometry \citep{2011ApJ...729..128C} of HR8799 b (upper limits for HR8799 c and d), and LBT/PISCES H-band of the four objects \citep{2012ApJ...753...14S}. We took the Keck Ks-band photometry of the four planets on top of the $CH_{4,S}/CH_{4,L}$ for HR8799cde published in \cite{2008Sci...322.1348M, 2010Natur.468.1080M}. The $CH_{4,S}/CH_{4,L}$ photometry of HR8799 b and c is in agreement with the P1640 and OSIRIS data. The  $CH_{4,L}$ photometry of HR8799c however does not agree with the SPHERE H3 photometry although the two filters overlap. It is at odds with the current knowledge of the planet SED and was discarded (see Section \ref{subsec:empirical}). We also chose not to account for the HST photometry of HR8799b obtained by \cite{2015arXiv150802395R} because of the discrepancy of this photometry with that obtained at $z$ and J bands from the ground and whose origin is unclear (see Section \ref{subsec:var}). The HST photometry obtained for the c planet is consistent, but redundant, with the P1640 spectrum. We did not consider it as well for that reason.
 
Longward of 2.5 $\mu$m, we added the Keck/NIRC2 L' and VLT/NaCo $[4.05]$  photometry from \cite{2014ApJ...795..133C}, and the M' band photometry from \cite{galicher11}. We considered the $[3.3 \mu m]$ photometry for the four planets and the narrow-band (L\_ND1 to L\_ND5) photometry of HR8799c and d obtained  with LBT/LMIRCam  by \cite{2012ApJ...753...14S} and \cite{2014ApJ...792...17S}.

 We converted each published contrast to fluxes using the flux-calibrated model spectra of HR8799A produced in paper III.  We summarized the fluxes of HR\,8799\,bcde  considered for this study into Tables~\ref{TabApA:HR8799b}, \ref{TabApA:HR8799c}, \ref{TabApA:HR8799d}, and \ref{TabApA:HR8799e}.  They are plotted in Figure \ref{Fig:SEDemp}.
 
The following analysis assumes that if the planets turn out to be variable, the multi-epoch measurements considered above are not strongly affected by the variability. 
 
\begin{table*}[t]
\begin{minipage}[ht]{\linewidth}
\caption{Currently available  fluxes of HR\,8799\,b at 10~pc gathered from narrow-band and broad-band photometry.}
\label{TabApA:HR8799b}
\begin{center}
\renewcommand{\footnoterule}{}  
\begin{tabular}{lllllll}
\hline \hline 
Filter &$\lambda$ &FWHM &Abs. Flux & err. Flux + & err. Flux - & Ref \\
  &($\mu$m) &($\mu$m) &(W.m-2.$\mu$m$^{-1}$) &(W.m-2.$\mu$m$^{-1}$) &(W.m-2.$\mu$m$^{-1}$) & \\
\hline 				
$z$- Subaru &1.02 &0.103 &2.99400e-16 &9.17000e-17 &7.02000e-17 & 2 \\
BB\_$J$ &1.245 &0.200 & 6.41999e-16 &6.72532e-17 & 6.08767e-17 & 1 \\
 $H2$ &1.593 &0.052 & 1.26405e-15 & 1.90374e-16 & 1.65454e-16 & 1  \\
 CH$_{4, S}$-Keck &1.5923 &0.1257 &1.02200e-15 &1.73000e-16 &1.48200e-16 & 3 \\
 $H$- LBT &1.633 &0.296 &1.10100e-15 &1.40000e-16 &1.24200e-16 & 4 \\
 $H3$ &1.667 & 0.054 & 1.41690e-15 & 1.60663e-16  & 1.44301e-16 &1 \\
CH$_{4, L}$- Keck &1.6809 &0.1368 &1.10600e-15 &1.99000e-16 &1.69100e-16 &3 \\
 He I B- Keck &2.0563 &0.0326 &6.02652e-16 &4.96300e-17 &4.96300e-17 &5 \\
 $K1$ &2.110 & 0.102 & 1.05039e-15  & 8.53835e-17  & 7.89649e-17 &1 \\
 $H_{2}$($\nu$=1-0) - Keck &2.1281 &0.0342 &1.00619e-15 &7.28200e-17 &7.28160e-17 &5 \\
 $K_{s}$- Keck &2.146 &0.311 &1.03700e-15 &8.00000e-17 &7.34000e-17 &3 \\
 Br$_{\gamma}$2 - Keck &2.1686 &0.0326 &1.05821e-15 &7.64899e-17 &7.65020e-17 &5 \\
 $K2$ & 2.251 &0.109 & 9.71028e-16 & 1.01722e-16 &  9.20764e-17 & 1 \\
 $H_{2}$($\nu$=2-1) - Keck &2.2622 &0.0388 &8.02563e-16 &6.44390e-17 &6.44390e-17 &5 \\
$[3.3]$ & 3.31 & 0.40 & 4.38455e-16 & 4.67493e-17 & 4.22455e-17 & 4 \\
 $L^{\prime}$- Keck &3.776 &0.700 &4.71200e-16 &4.54000e-17 &4.15000e-17 &6 \\
 $[4.05]$ &4.051 &0.02 &6.87300e-16 &1.24000e-16 &1.05000e-16 &6 \\
 $M^{\prime}$- Keck &4.670 &0.241 &1.28900e-16 &4.10000e-17 &3.11200e-17 & 7 \\
\hline
\end{tabular}
\end{center}
\end{minipage}
\tablefoot{[1] - Zurlo et al. 2015 (paper III), [2] - \cite{2011ApJ...729..128C}, [3] - \cite{2008Sci...322.1348M}, [4] - \cite{2012ApJ...753...14S}, [5] - \cite{2011ApJ...733...65B}, [6] - \cite{2014ApJ...795..133C}, [7] - \cite{galicher11}.}
 \end{table*}

\begin{table*}[t]
\begin{minipage}[ht]{\linewidth}
\caption{Same as Table \ref{TabApA:HR8799b} but for HR\,8799\,c.}
\label{TabApA:HR8799c}
\begin{center}
\renewcommand{\footnoterule}{}  
\begin{tabular}{lllllll}
\hline \hline 
Filter &$\lambda$ &FWHM &Abs. Flux & err. Flux + & err. Flux - & Ref \\
  &($\mu$m) &($\mu$m) &(W.m-2.$\mu$m$^{-1}$) &(W.m-2.$\mu$m$^{-1}$) &(W.m-2.$\mu$m$^{-1}$) & \\
\hline 				
 $z$- Subaru &1.02 &0.103 &<1.514e-15 &\dots &\dots & 2 \\
BB\_$J$ &1.245 &0.200 &1.90342e-15 & 2.68309e-16  & 2.35162e-16 &1 \\
 $H2$ &1.593 &0.052  & 3.14605e-15 & 4.13797e-16 &  3.65694e-16 &1 \\
 CH$_{4, S}$- Keck &1.5923 &0.1257 &2.407e-15 &4.60000e-16 &3.87000e-16 &3 \\
 $H$- LBT &1.633 &0.296 &2.52200e-15 &4.28000e-16 &3.65000e-16 &4 \\
  $H3$ & 1.667 & 0.054 & 3.55910e-15 & 4.03568e-16 & 3.62471e-16 &1 \\
 CH$_{4, L}$-Keck &1.6809 &0.1368 &2.75200e-15 &5.27000e-16 &4.42000e-16 &3 \\
  $K1$ &2.110 & 0.102 & 2.54302e-15 & 1.89672e-16 & 1.76507e-16&1 \\
 $K_{s}$- Keck &2.146 &0.311 &2.42000e-15 &1.86000e-16 &1.71000e-16 &3 \\
 $K2$ & 2.251 &0.109 & 2.69918e-15 & 2.39213e-16 & 2.19739e-16  &1 \\
 L\_ND1 &3.04 &0.15 &1.34456e-15 &1.99199e-16 &1.73497e-16 &5 \\
 L\_ND2 &3.16 &0.08 &1.12583e-15 &2.27715e-16 &1.89405e-16 &5 \\
$ [3.3]$ &3.31 &0.40 &1.10135e-15 &1.17429e-16 &1.06116e-16 &4 \\
 L\_ND3 &3.31 &0.16 &1.12904e-15 &1.67271e-16 &1.45687e-16 &5 \\
 L\_ND4 &3.46 &0.16 &1.44899e-15 &2.14673e-16 &1.86971e-16 &5 \\
 L\_ND5 &3.59 &0.06 &1.50659e-15 &2.23207e-16 &1.94403e-16 &5 \\
 $L^{\prime}$- Keck &3.776 &0.700 &1.04000e-15 &8.00000e-17 &7.35000e-17 &6 \\
 L\_ND6 &3.78 &0.19 &1.22305e-15 &1.81199e-16 &1.57816e-16 &5 \\
 $[4.05]$ &4.051 &0.02 &1.50400e-15 &1.15000e-16 &1.07000e-16 &6 \\
 $M^{\prime}$- Keck &4.670 &0.241 &3.29800e-16 &4.54000e-17 &3.99000e-17 &7 \\
\hline
\end{tabular}
\end{center}
\end{minipage}
\tablefoot{[1] - Zurlo et al. 2015 (paper III), [2] - \cite{2011ApJ...729..128C}, [3] - \cite{2008Sci...322.1348M}, [4] - \cite{2012ApJ...753...14S}, [5] - \cite{2014ApJ...792...17S}, [6] - \cite{2014ApJ...795..133C}, [7] - \cite{galicher11}.}
 \end{table*}

\begin{table*}[t]
\begin{minipage}[ht]{\linewidth}
\caption{Same as Table \ref{TabApA:HR8799b} but for HR\,8799\,d.}
\label{TabApA:HR8799d}
\begin{center}
\renewcommand{\footnoterule}{}  
\begin{tabular}{lllllll}
\hline \hline 
Filter &$\lambda$ &FWHM &Abs. Flux & err. Flux + & err. Flux - & Ref \\
  &($\mu$m) &($\mu$m) &(W.m-2.$\mu$m$^{-1}$) &(W.m-2.$\mu$m$^{-1}$) &(W.m-2.$\mu$m$^{-1}$) & \\
\hline 				
 $z$- Subaru &1.02 &0.103 &<5.758e-15 &\dots &\dots &2 \\
BB\_$J$ &1.245 &0.200 &1.92103e-15 & 7.92081e-16 & 5.60835e-16 &1 \\
 $H2$ &1.593 &0.052  & 3.35557e-15 &6.06092e-16 & 5.13367e-16 &1 \\
 CH$_{4, S}$-Keck &1.5923 &0.1257 &2.94700e-15 &9.38000e-16 &7.11000e-16 &3 \\
 $H$- LBT &1.633 &0.296 &2.40900e-15 &5.41000e-16 &4.42000e-16 &4 \\
$H3$ & 1.667 & 0.054 & 3.33687e-15  & 5.68760e-16 & 4.85937e-16 &1 \\
 CH$_{4, L}$- Keck &1.6809 &0.1368 &1.48500e-15 &3.50000e-16 &2.84000e-16 &3 \\
  $K1$ &2.110 & 0.102 &2.51970e-15  & 2.23310e-16 & 2.05128e-16 &1 \\
 $K_{s}$- Keck &2.146 &0.311 &2.46500e-15 &2.89000e-16 &2.58000e-16 &3 \\
 $K2$ & 2.251 &0.109 & 2.74936e-15 & 3.11751e-16 & 2.80002e-16  &1 \\
 L\_ND1 &3.04 &0.15 &1.34456e-15 &2.71950e-16 &2.26210e-16 &5 \\
 L\_ND2 &3.16 &0.08 &1.29263e-15 &3.34690e-16 &2.65860e-16 &5 \\
 $[3.3]$ &3.31 &0.40 &1.32411e-15 &1.41180e-16 &1.27580e-16 &4 \\
 L\_ND3 &3.31 &0.16 &9.83352e-16 &1.98898e-16 &1.65435e-16 &5 \\
 L\_ND4 &3.46 &0.16 &1.44899e-15 &2.93080e-16 &2.43770e-16 &5 \\
 L\_ND5 &3.59 &0.06 &1.57760e-15 &2.33720e-16 &2.03570e-16 &5 \\
 $L^{\prime}$- Keck &3.776 &0.700 &1.20600e-15 &1.04000e-16 &9.60000e-17 &6 \\
 L\_ND6 &3.78 &0.19 &1.28069e-15 &1.89730e-16 &1.65260e-16 &5 \\
 $[4.05]$ &4.051 &0.02 &1.64900e-15 &2.27000e-16 &2.00000e-16 &6 \\
 $M^{\prime}$- Keck &4.670 &0.241 &4.68000e-16 &1.78000e-16 &1.29000e-16 &7 \\
\hline
\end{tabular}
\end{center}
\end{minipage}
\tablefoot{[1] - Zurlo et al. 2015 (paper III), [2] - \cite{2011ApJ...729..128C}, [3] - \cite{2008Sci...322.1348M}, [4] - \cite{2012ApJ...753...14S}, [5] - \cite{2014ApJ...792...17S}, [6] - \cite{2014ApJ...795..133C}, [7] - \cite{galicher11}.}
 \end{table*}

\begin{table*}[t]
\begin{minipage}[ht]{\linewidth}
\caption{Same as Table \ref{TabApA:HR8799b} but for HR\,8799\,e.}
\label{TabApA:HR8799e}
\begin{center}
\renewcommand{\footnoterule}{}  
\begin{tabular}{lllllll}
\hline \hline 
Filter &$\lambda$ &FWHM &Abs. Flux & err. Flux + & err. Flux - & Ref \\
  &($\mu$m) &($\mu$m) &(W.m-2.$\mu$m$^{-1}$) &(W.m-2.$\mu$m$^{-1}$) &(W.m-2.$\mu$m$^{-1}$) & \\
\hline 				
 BB\_$J$ &1.245 &0.200 &2.28841e-15  & 5.09894e-16  & 4.16985e-16 &1 \\
 $H2$ &1.593 &0.052  & 3.71335e-15  & 7.87435e-16 & 6.49673e-16 &1 \\
 $H3$ & 1.667 & 0.054 &3.90247e-15  & 8.69535e-16 & 7.11092e-16 &1 \\
 $H$- LBT &1.633 &0.296 &3.32500e-15 &7.47000e-16 &6.10000e-16 &3 \\
 $K1$ &2.110 & 0.102  & 2.71237e-15 & 3.07557e-16  & 2.76235e-16 &1 \\
 $K2$ & 2.251 &0.109 &2.80048e-15 & 3.42574e-16 &  3.05238e-16 &1 \\
 $K_{s}$- Keck &2.146 &0.311 &3.01900e-15 &8.17000e-16 &6.43000e-16 &2 \\
 $[3.3]$ &3.31 &0.40 &1.32411e-15 &2.82550e-16 &2.32860e-16 &3 \\
 $L^{\prime}$- Keck &3.776 &0.700 &1.21700e-15 &1.42000e-16 &1.28000e-16 &4 \\
 $[4.05]$ &4.051 &0.02 &1.89300e-15 &3.83000e-16 &3.18000e-16 &4 \\
 $M^{\prime}$- Keck &4.670 &0.241 &<2.006e-15 &\dots &\dots &5 \\
\hline
\end{tabular}
\end{center}
\end{minipage}
\tablefoot{[1] - Zurlo et al. 2015 (paper III), [2] - \cite{2010Natur.468.1080M}, [3] - \cite{2012ApJ...753...14S}, [4] - \cite{2014ApJ...795..133C}, [5] - \cite{galicher11}.}
 \end{table*}

\section{Empirical study}
\label{subsec:empirical}
\subsection{Comparison to benchmark objects}
We showed in paper III that among a large sample of MLT field dwarfs  from the SpeXPrism library, the colors of HR8799bcde and spectra of the d and e planets were best reproduced by those of some dusty L dwarfs. We took advantage of the new SPHERE data to make a  more in-depth empirical analysis of the planet spectra and photometric data points. 

\subsubsection{Colors}

\begin{figure}[!]
\begin{center}
\includegraphics[width=\columnwidth]{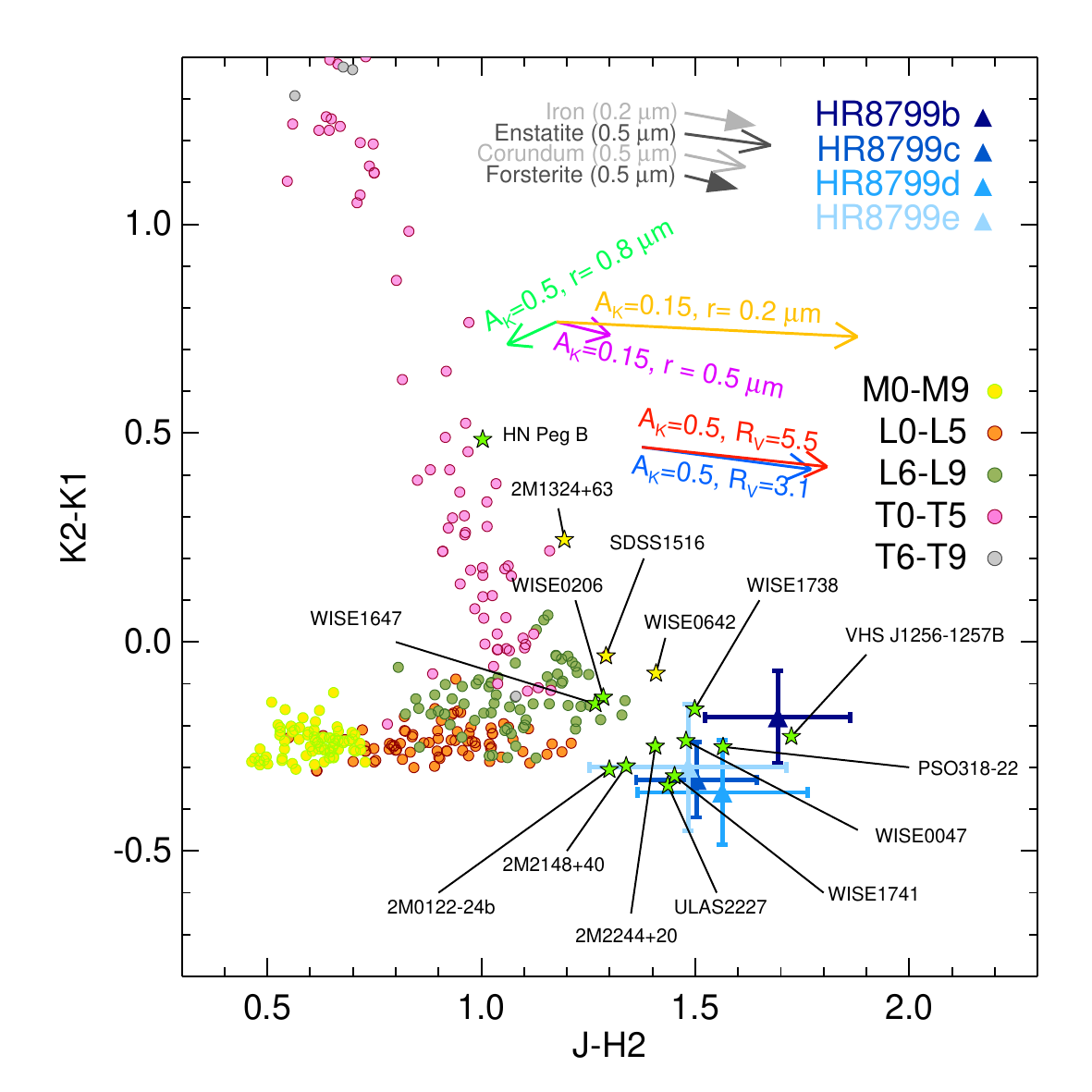}
\caption{Comparison of HR\,8799\,bcde colors inferred from the IRDIS photometry of the system to those of M, L, T field dwarfs (dots), and of young companions and red dwarfs straddling the L/T transition (green stars). Reddening vectors for a 0.5 mag $K$ band extinction and reddening parameters $R_{V}=3.1$ and 5.5 are overlaid (red and blue arrows). We  show the reddening vector for forsterite dust grains with size distribution centered on radii $r=0.8~\mu$m (green arrow), $r=0.4~\mu$m (violet-purple arrow), and $r=0.2~\mu$m (orange arrow), as well as the vectors for different grain species in gray ($A_{K}=0.15$ mag considered).}
\label{f:JH2K2K1}
\end{center}
\end{figure}

\begin{figure}[!]
\begin{center}
\includegraphics[width=\columnwidth]{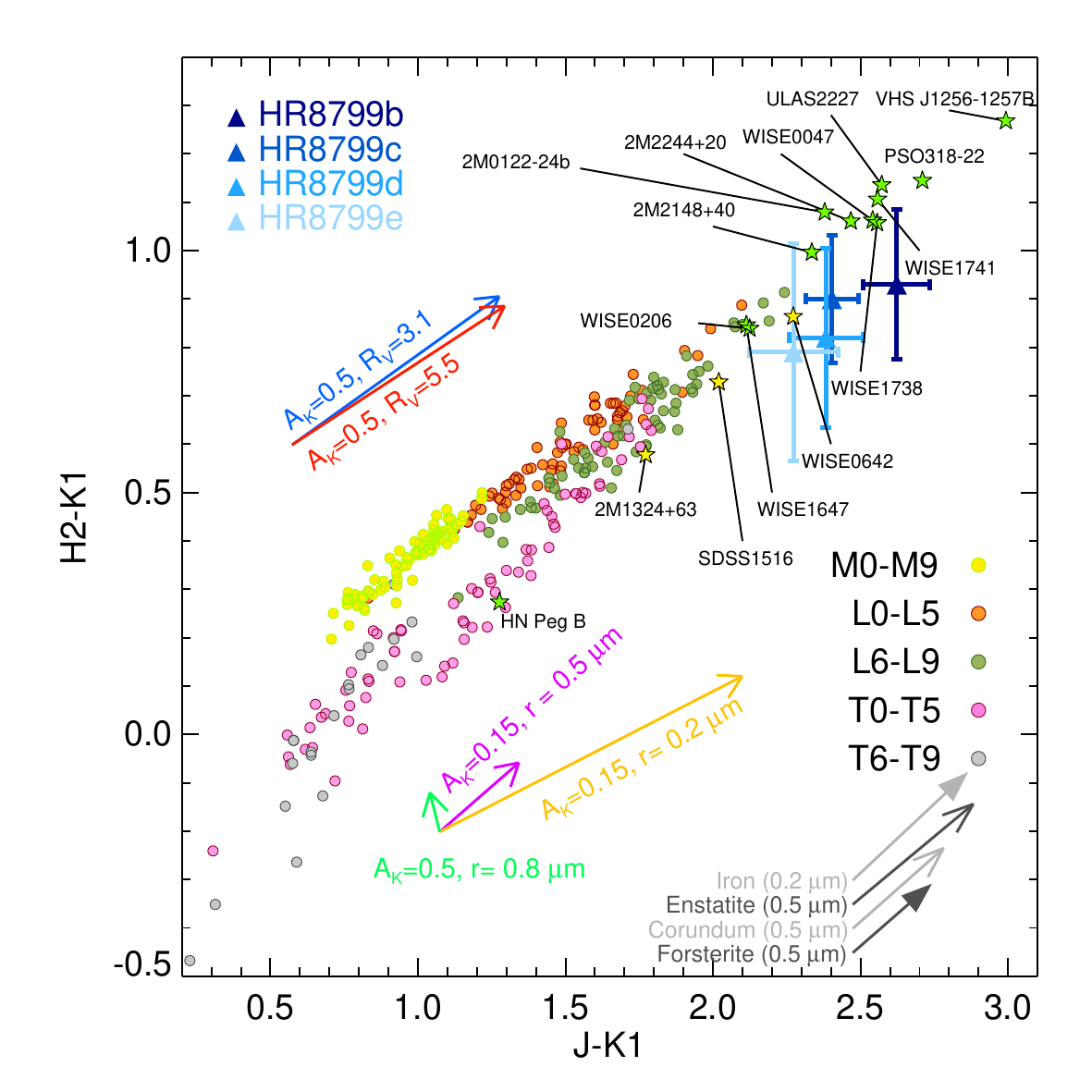}
\caption{Same as Fig.~\ref{f:JH2K2K1} but for the $J-K$1 versus $H2-K1$ colors.}
\label{f:JK1H2K1}
\end{center}
\end{figure}

We use the two  color-color diagrams shown in paper III  and added the colors of the red L and T dwarfs (green and yellow stars in Figures~\ref{f:JH2K2K1} and \ref{f:JK1H2K1}): \object{WISEPA J020625.26+264023.6} and \object{WISEPA J164715.59+563208.2} \citep[L9pec,][]{2011ApJS..197...19K}, \object{PSO J318.\-53\-38-22.86\-03} \citep[L$7\pm1$,][]{2013ApJ...777L..20L}, \object{WISE J174102.78-464225.5} \citep[L7pec $\pm2$,][]{2014AJ....147...34S}, \object{WISEP J004701.06+680352} \citep[L7pec,][]{2015ApJ...799..203G},  \object{ULAS J222711-004547} \citep[L7pec,][]{2014MNRAS.439..372M}, \object{WISEJ064205.58+410155.5} \citep[L/Tpec,][]{2013ApJS..205....6M}, \object{WISE J1738\-59.27+61\-4242.1} \citep[L/Tpec,][]{2013ApJS..205....6M}.  For that purpose, we smoothed the low-resolution spectra of the objects to the resolution of the narrowest IRDIS filter, used the IRDIS filter pass-bands, a model of the Paranal atmospheric transmission generated with the ESO \texttt{Skycalc} web application\footnote{http://www.eso.org/obser\-ving/etc/bin/gen/form?INS.MO\-DE=swspectr+INS.NA\-ME=SKYCALC} \citep{2012A&A...543A..92N, 2013A&A...560A..91J}, and a model spectrum of Vega \citep{2007ASPC..364..315B}. The same procedure was applied to the continuous 1.1-2.5~$\mu$m spectra of the moderately young companions \object{HN Peg B} \citep[age $\sim0.3\pm0.2$ Gyr,][]{2007ApJ...654..570L}, \object{VHS 1256-1257 B} \citep{2015ApJ...804...96G}, the younger companion \object{2MASS J01225093-2439505 b}  \citep[possibly $\sim$ 10-120 Myr old,][Rojo et al. 2015, in prep.]{2013ApJ...774...55B}.

In the color-color diagrams, the red L6-L8 dwarfs and the young and dusty L4-L6 companions 2MASS J01225093-2439505\,b and VHS J125601.92-125723.9 B deviate from the sequence of field dwarfs. These objects have similar colors  to HR\,8799\,bcde planets.  The  dwarfs WISE0206, WISE0642, and WISE1647 together with the young T2 companion HN\,Peg\,B and the T0.5-T1.5 dwarf \object{SDSS J151643.01+305344.4} (hereafter SDSS J1516) seem to be later-type objects than the L6-L8 red dwarfs, but still have redder colors than the sequence of field dwarfs. 

\subsubsection{Spectra and SEDs}

\begin{figure}
\begin{center}
\includegraphics[width=9.4cm]{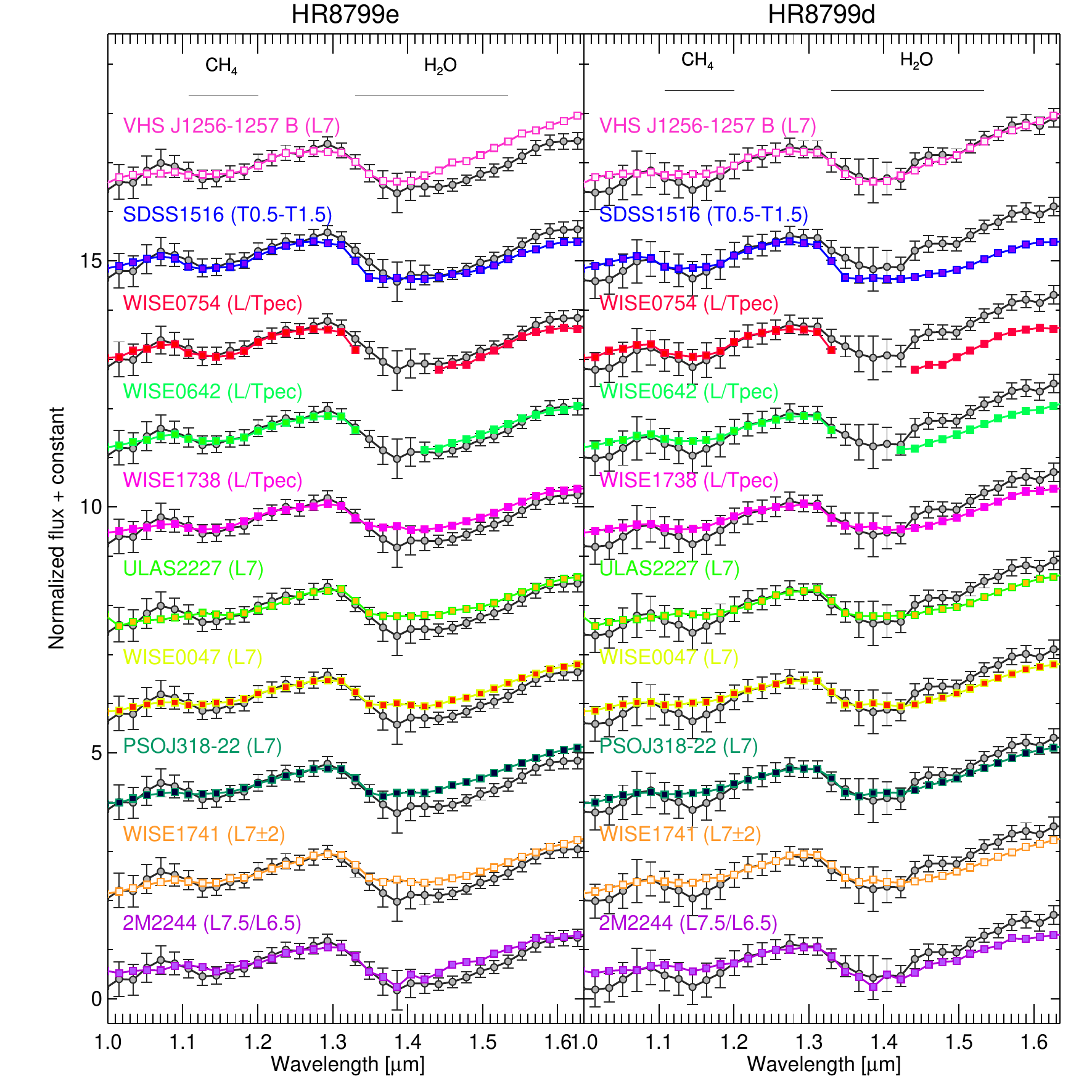}
\caption{Comparison of HR\,8799\,d and e normalized low-resolution SPHERE YJH band spectra to those of red late-L and early-T dwarfs.}
\label{Fig:youngdusty}
\end{center}
\end{figure}

We compare in Fig.~\ref{Fig:youngdusty} the IFS spectra of HR\,8799\,d and e to the ones of the aforementioned red dwarfs and companions. We considered as well a comparison to the spectra of the companions \object{HD 203030B} \citep[age $\sim$ 0.13-0.4 Gyr,][Rojo et al. 2015, in prep]{2006ApJ...651.1166M} and 2M1207b \citep[8 Myr,][]{2010A&A...517A..76P}. This enables to identify 1.1-1.2 $\mu$m and $1.35-1.55\mu$m $H_{2}O$ absorption of both  exoplanets. The HR\,8799\,d spectrum is best reproduced by the one of the young L7 companion VHS J125601.92-125723.9 b. The one of HR\,8799\,e is also well fitted by the one of the peculiar L/T transition dwarf WISE0642.

Ultimately, we attempted a comparison of the 0.96 to 4.67~$\mu$m SED of the planets to those of red L dwarfs built from their flux-calibrated near-infrared spectra and WISE $W1$ ($\lambda_{c}=3.35~\mu m$, bandwidth=$0.66~\mu m$) and $W2$ ($\lambda_{c}=4.60~\mu m$, bandwidth=$1.04~\mu m$) photometry, or taken from \cite{2009ApJ...702..154S}.  

\begin{figure*}
\begin{center}
\includegraphics[width=13cm]{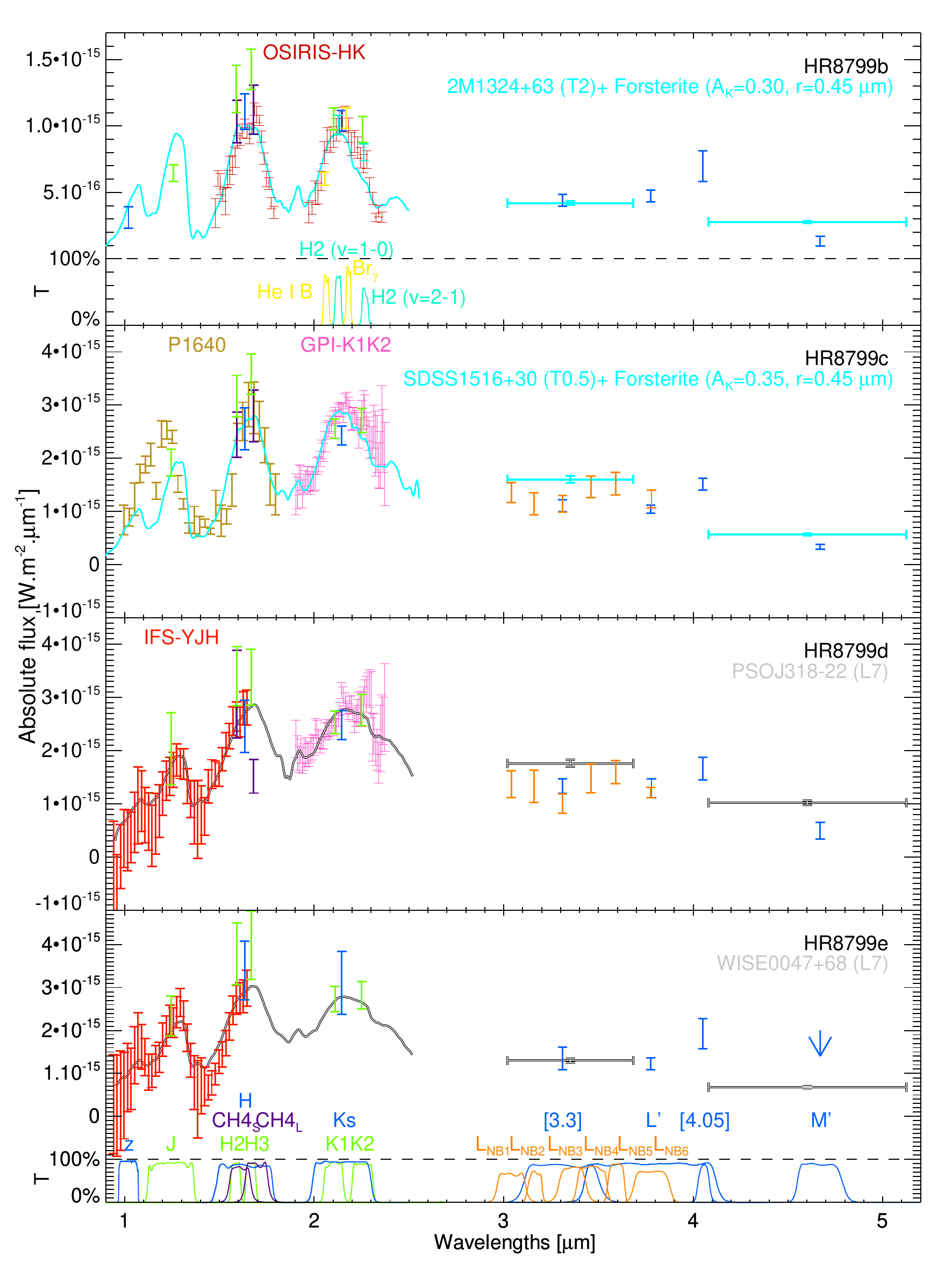}
\caption{Comparison of the current set of spectrophotometric points of HR\,8799\,b,c,d, and e to the normalized SEDs of red L dwarfs (gray) and to the SED of peculiar early-T dwarfs reddened by additional forsterite extinction (light blue).}
\label{Fig:SEDemp}
\end{center}
\end{figure*}

The results are reported in Fig.~\ref{Fig:SEDemp}. Only a few known objects match the SED of the d and e planets. The SED of HR\,8799\,e is well reproduced by the one of WISE0047 (L7). The two red dwarfs WISE1738 (late-L) and WISE 1741 (L7) also provide a decent fit, but reproduce less well the IFS spectrum of the planet. The SED of HR8799d is best fitted by the one of PSOJ318-22 (L7). We find that the two L7 dwarfs WISE1741 and ULAS2227 provide a better match of the 3-4 $\mu$m flux of the planet at the price of a worse fit of the IFS spectra.

We do not find a good fitting template for HR8799b and c. The L/T transition dwarf WISE1741-46 reproduces the presently available SED of HR\,8799\,c, apart for the strength of the water absorption from 1.9 to 2.0~$\mu$m.  ULAS2227 and WISE1741 provide a good fit of the $Y$, $J$, $[3.3]$, $L^{\prime}$, and $M^{\prime}$ band fluxes for the planet b but fail to reproduce the shape of the OSIRIS $HK$ band spectrum obtained by \cite{2011ApJ...733...65B}. The deeper water-bands in the $H$ and $K$ bands (see Fig.~\ref{Fig:SEDemp}) in the spectrum of these planets suggest that they both have latter spectral types than HR\,8799\,d and e. The comparison to newly discovered L9pec-T2pec dwarfs in the near-future \citep[e.g.][]{2014AAS...22344121K} may nevertheless solve the issue. The good matches found from the analysis of the SED and spectra of HR8799d and e confirm the analysis based on Figures~\ref{f:JH2K2K1} and \ref{f:JK1H2K1}.  \\

Most of the companions and red dwarfs representing well the properties of HR8799d and e are thought to be low-mass and low surface gravity objects in the Solar neighbourhood.  \cite{2014ApJ...783..121G} deduce that WISE0047 and PSOJ318-22 are strong candidates of the AB\,Doradus and $\beta$\,Pictoris moving groups, respectively. WISE1741 is also proposed as potential member of one of these two groups \citep{2014AJ....147...34S}. We note that, if truly members of these groups, all these objects have estimated masses 6-18~\MJup, which bracket the present mass estimates of the two planets. 

\subsection{Effect of additional cloud opacity}
The deviation of the colors of young and dusty dwarfs, and of the HR\,8799 planets with respect to the sequence of field dwarfs in Fig.~\ref{f:JH2K2K1} and \ref{f:JK1H2K1} follows  the reddening vector computed from the synthetic interstellar extinction curves of \citet{2003ARA&A..41..241D}, \citet{2003ApJ...598.1017D}, and \citet{2003ApJ...598.1026D} with $R_{V}=3.1$ and 5.5. These curves are based on a model of extinction caused by a mixture of silicate and carboneous grains with log-normal size distributions extending from 3.5 $\AA$ to $\mu m$-size\footnote{The smallest carboneous grains are PAH molecules.  The largest grains consist of graphite. And grains of intermediate size have optical properties intermediate between those of PAHs and graphite. The silicate grains are composed of olivine species \citep{1984ApJ...285...89D}.}. The models with $R_{V}=3.1$ and 5.5 consider grain abundances per unit H column of C/H=  55.8 ppm (in log-normal size distributions) and 42.6 ppm respectively.  This result is consistent with the results of \citet{2014MNRAS.439..372M}, who show that the spectra of red L6-L8 dwarfs could match partially the 0.7-2.5~$\mu$m spectra of mid to late-L field-dwarfs once dereddened by interstellar extinction laws \citep{1989ApJ...345..245C, 1999PASP..111...63F}. 

  \citet{2014MNRAS.439..372M} showed that the near-infrared spectra of red L dwarfs dereddened by corundum (Al$_{2}$O$_{3}$), iron  (Fe), or enstatite (MgSiO$_{3}$) extinctions could also provide a good match to standard field dwarfs spectra.  These three grains species along with forsterite (Mg$_{2}$SiO$_{4}$) are expected to play a dominant role onto the chemistry and radiative transfer into the atmosphere of brown dwarfs and warm young giant planets \citep{1996ApJ...472L..37F, 1996A&A...308L..29T, 2001ApJ...556..357A, 2002Icar..155..393L, 2010ApJ...716.1060V}.    Therefore, we used the extinctions curves of corundum, enstatite, and iron computed by \cite{2014MNRAS.439..372M} for a range of characteristic  grain radii (hereafter $r$) from 0.05 to 1.00 $\mu$m, in steps of 0.05 $\mu$m for the analysis of HR8799 planet properties. For each mean grain size, a Gaussian size distribution of width $\sqrt{2\sigma}=0.1 \times r$ is adopted. We also computed for this work  the extinction curves of  amorphous forsterite (Mg$_{2}$SiO$_{4}$) following the procedure described in \cite{2014MNRAS.439..372M}. We used for  those grains the optical constants of  \cite{1996ApJS..105..401S}. Corundum has a higher condensation temperature than the other three species considered here \citep{2003ApJ...591.1220L}. Consequently, it is expected to be found into the hotter deeper atmospheric layers, unless its distribution is affected by (vertical) mixing. This grains is also expected to occur in an order of magnitude lower concentrations than forsterite and enstatite in a solar composition (because Al is  at least 10 times less abundant than Mg and Si).  

We used the extinction curves of the grains species considered here to compute reddening vectors for the SPHERE colors. We report them in Fig.~\ref{f:JH2K2K1} and \ref{f:JK1H2K1}. The reddening vectors for the four grain types vary considerably in direction and norm with the grain size, but they display qualitatively the same behavior. This is illustrated in Fig.~\ref{f:JH2K2K1} and \ref{f:JK1H2K1} for the forsterite. We also compare the reddening vectors for the four grain species  (gray arrows) for a $A_{K}$=0.15 mag. The orientation of the vectors for the forsterite, corundum, and the enstatite   reproduces  well the color deviation of the HR\,8799 planets and of the red L dwarfs only for grains with sizes  distributed around $r \sim 0.3-0.6~\mu$m. The iron grains must have sizes distributed around 0.2 $\mu$m to reproduce the color deviation in both diagrams.   This is consistent with the best dereddening values found by \cite{2014MNRAS.439..372M} for most of the red L6-L7 dwarfs when compared to the spectra of L4.5-L7 field dwarfs. The reddening vectors found for the 0.5 $\mu$m forsterite grains are also consistent with the reddening vectors produced by interstellar  dust, which consider this grain species among a mixture of silicates.

Given the lack of proper empirical templates for HR\,8799\,b and c, we investigated whether the available photometry for the two planets could be reproduced by the spectrum of a dwarf later than L7 and reddened by the grain opacities studied here. We considered as original templates the spectra of \cite{2009ApJ...702..154S} which cover the 3-4~$\mu$m region as well as spectra of standard L and T fields dwarfs from the SpeXPrim library ($\sim$0.65-2.55 $\mu$m) combined with the existing WISE photometry of the sources. We let the 2MASS $K_{S}$ band extinction varies between 0 and 3.0 mag in steps of 0.05 mag as well as the grain size (see above) in the fitting process.We used the $g$\as goodness-of-fit indicator defined by \cite{2010ApJ...723..850B} to identify the best-fitting template in the wavelength interval 0.95-2.5 $\mu$m.  This weighted $\chi^{2}$ indicator accounts for an inhomogeneous sampling of the SED of the object and errors bars on the template and the object. We took the maximum uncertainty of our assymetric error bars to compute $g$\as since we did not have apriori on the posterior distributions for the photometric values. The fit was repeated for the four grain species considered above. To conclude, we determined visually for each grain species which combination of $A_{K}$ and grain size  was needed to ajust the WISE W1 and W2 photometry of the templates \citep{2012yCat.2311....0C} together with their 1-2.5 $\mu$m fluxes onto the SED of the planets.

Among L and T-type field dwarfs standards\footnote{from the SPeXPrism library \citep{2006AJ....131.1007B, 2007ApJ...659..655B, 2010ApJ...710.1142B, 2004AJ....127.2856B, 2006ApJ...639.1095B, 2007AJ....134.1162L}}, only T1-T2 objects reddenned either by the extinctions of forsterite, enstatite, corundum, or iron reproduce the slope of the SED of HR8799b and c. However, the reddened templates fail to represent the shape and flux of the planets in the $K$ band. These differences arise from a variation of the collision-induced absorption of $H_{2}$ that tends to suppress the $K$ band flux in higher surface-gravity atmospheres \citep{1997A&A...324..185B}.

The missmatch in the K band is reduced  for HR8799b considering reddenned spectra of the peculiar T2 brown dwarf \object{2MASS J13243553+6358281} \citep{2007AJ....134.1162L,2008ApJ...676.1281M}.  Conversely, we find that  the  reddened spectrum of  \object{SDSS J1516} represents well the 1-2.5 $\mu$m flux od HR8799c. It has been proposed that these sources are  younger than the field population \citep[$<$ 300~Myr,][]{2007AJ....134.1162L, 2009ApJ...702..154S}. \cite{2012ApJ...754..135M} noticed that SDSS J1516 has similar cloud properties as HR8799b,c, and d  based on atmospheric modelling. The reddened spectrum of the extremely red dwarf \object{WISEJ064205.58+410155.5} \citep{2013ApJS..205....6M}, which could fall into the T-dwarf category, provides a good alternative fitting solution to the SED of HR\,8799\,c in the 0.95-2.5 $\mu$m range. But its dereddened WISE photometry fails to reproduce the planet flux longward of 3 $\mu$m. As noted by \cite{2013ApJS..205....6M}, this object lacks the CH$_{4}$ absorption in its spectrum, as HR\,8799\,c. \cite{2014ApJ...783..121G} propose it as a candidate of the AB\,Doradus moving group with a modest probability (53\%). Similarly, the reddened spectrum  of the extremely red T dwarf \object{WISEJ075430.95+790957.8}  \citep{2013ApJS..205....6M} reproduces well the $H$ band shape of HR\,8799\,b, especially the red slope/plateau from 1.6 to 1.72 $\mu$m and the lack of a methane absorption longward of 1.6 $\mu$m. But the fit is not as good for the rest of the SED. The disagreement between the objects SEDs longward of 3~$\mu$m could be due to a blend of its WISE photometry with that of a neighboring source \citep{2013ApJS..205....6M}. 

We summarize into Table \ref{Tab:der} the results of the fits of the 1-2.5 $\mu$m and 1-5 $\mu$m SED of the planets considering these four peculiar objects as templates and the different grain species. The iron grains used for the extra reddenning need to be of smaller sizes than those of forsterite, enstatite, and corundum. This is in agreement with \cite{2014MNRAS.439..372M} and the reddening vectors shown in Figures \ref{f:JH2K2K1} and \ref{f:JK1H2K1}. We  show in Figure \ref{Fig:SEDemp} the best fit of the whole SED of HR8799b and c (all grain species considered, all empirical templates considered). For each grain specie, we always find a combination of $A_{K}$ and $r$ which reproduces the SED of the planets. The forsterite grain sizes needed (0.45-0.70 $\mu$m) for SDSS J1516 and \object{WISEJ064205.58+410155.5} to match the whole SED of HR8799c are roughtly consistent with the reddenning vectors  needed to bring the IRDIS colors of these objects to the location of the planet c in Figures \ref{f:JH2K2K1} and \ref{f:JK1H2K1} (yellow stars). The same is true for  \object{2MASS J13243553+6358281} with respect to HR8799 b. We did not compare the position of WISEJ075430.95+790957.8 with respect to HR8799 b in these diagrams since the synthetic J-band IRDIS magnitude is biased by artefacts in the spectrum of this source around 1.37 $\mu$m.  However, this object seems to requiere a reddenning with bigger forsterite, enstatite, or corundum grains than  2MASS J13243553+6358281 to reproduce the SED of HR8799b. \\

In summary, our empirical modelling of the presently available spectrophotometry of HR\,8799\,d and e suggests that these planets are part of a growing population of dusty L6-L8 dwarfs. Furthermore, we show that the properties of the planets b and c can be explained empirically by an increased amount of occulting photospheric dust with respect to the photosphere of red -- and possibly young --  early-T dwarfs known to date. The empirical analysis of HR\,8799\,b and c is clearly limited by the lack of young and dusty free-floating objects with spectral types L9-T2. This motivates the use of atmospheric models to complement the understanding of the planet properties. 

\begin{table*}[t]
\centering
\caption{\label{Tab:der}Reddening parameters needed to ajust the peculiar T-type templates onto the 1-2.5 $\mu$m spectrophotometry of HR8799b and c ($r$, 	$A_{K}$) and to represent the 1--5 $\mu$m SED of these two planets ( $r_{SED}$,  $A_{K_{SED}}$).}
\begin{tabular}{llllllll}
\hline
Planet	&	Template	& Grain specie	&	$r$	&	$A_{K}$ & g'' & $r_{SED}$ & $A_{K_{SED}}$ \\
			&						&						& ($\mu$m) & (mag) &   \\
\hline
\hline
b			&		2MASSJ13243553+6358281		&		Corundum	& 0.45	&	0.30	&	1.93	& 0.40	&	0.50	\\
			&														&		Enstatite	& 0.45	&	0.20	&	1.99	& 0.50	&	0.40  \\
			&														&		Forsterite	& 0.45   &  0.30	&	1.87  & 0.45	&	0.30	\\	
			&														&		Iron			& 0.20	&	0.30	&	2.27  &	0.20	&	0.40	\\	
\hline
b			&		WISEJ075430.95+790957.8		&		Corundum	& 0.70	&	0.80	&	2.90	 & 0.75	&	2.00 \\
			&														&		Enstatite	& 0.80	&	0.80	&	3.15	& 0.80	&	2.00	\\
			&														&		Forsterite	& 0.65   &  0.90   & 2.95    & 0.70	&	2.20  \\
			&														&		Iron			&	0.25  &  0.60	&	2.23 & 0.30	&	2.40  \\
\hline			
c			&		SDSSJ151643.01+305344.4		&		Corundum	&	0.70	&	1.20	&	6.71	& 0.45	&	0.35 \\
			&														&		Enstatite	&	0.80	&	1.40	&	6.75	 & 0.55	&	0.40	\\
			&														&		Forsterite	&	0.65	&	1.40	&	6.60  & 0.45	&	0.35	\\
			&														&		Iron			&	0.25	&	0.90	&	6.62  & 0.20	&	0.35	\\
\hline
c			&		WISEJ064205.58+410155.5		&		Corundum	&	0.80	&	2.0	&	6.28	&	0.55	&	0.45	\\
			&														&		Enstatite	&	0.85	&	1.4	&	6.43	&	0.40	&	0.60	\\
			&														&		Forsterite	&	0.70	&	1.4	&	6.55  &	0.50	&	0.40	\\
			&														&		Iron			&	0.25	&	0.7	&	7.60  &	0.20	&	0.25	\\
\hline
\end{tabular}
\end{table*}

\section{Spectral synthesis}
\label{subsec:model}
\subsection{Description of the models and fitting procedure}
We compared the available spectrophotometry of  HR\,8799\,bcde to the predictions of three atmospheric models:

\begin{itemize} 
\item The Exo-REM models  \citep{Baudino15}.  The models account for the formation of condensates from Si and Fe, the two
most abundant condensing elements in exoplanets. They do not account for non-equilibrium chemistry. But they are the only models used here to consider the most recent (EXOMOL) $CH_{4}$ linelist \citep{2014MNRAS.440.1649Y}. In addition of $T\mathrm{_{eff}}$, log $g$, and M/H, the models require as input the optical depth of the iron cloud ($\tau_{ref}$) at 1.2 $\mu$m and the mean radius $r$ of the particle size.  The first  grid of synthetic spectra (hereafter Exo-REM1) corresponds to models where no cloud forms in the atmosphere of the object ($r=0 \mu$m, $\tau_{ref}$=0.). The second grid (hereafter Exo-REM2) considers $r=30 \mu$m and $\tau_{ref}=0.1$ and corresponds to atmospheres with thin clouds.  The third (Exo-REM3) and fourth grids (Exo-REM4) correspond to models with the same particle sizes but  $\tau_{ref}$=1 and 3. The Exo-REM4 models correspond to the case of a strong impact of the cloud cover ("thick clouds"). To conclude, the fifth grid (Exo-REM5)  explores the case of an atmosphere with smaller dust grains ($r=3 \mu$m) and a medium content in dust ( $\tau_{ref}$=1). The relation between $\tau_{ref}$ and the optical depth of iron and forsterite at a pressure level of one bar is given in Table 3 of \cite{Baudino15} for the five flavors of Exo-REM models described above.

\item The parametric models of \cite{2011ApJ...737...34M} already used to represent the SED of HR8799bcde \citep{2011ApJ...737...34M, 2014ApJ...795..133C}. We considered the "thick cloud" AE   models where the particle density decays at the cloud tops with a scale height that is equal to twice that of the gas. The cloud particles are distributed in size using the Deirmendjian distribution \citep{2000ApJ...538..885S}. We used the models with a mean particle size of 60$\mu$m since the available grids covered a large interval in log $g$ and $T\mathrm{_{eff}}$. The models do not handle non-equilibrium chemistry.

\item The 2014 version of the \texttt{PHOENIX} BT-Settl atmospheric models.
The models are described in \cite{2014IAUS..299..271A} and \cite{2015A&A...577A..42B}. Recent updates to the brown dwarf and planetary domain are reported in
Vigan et al., submitted. They consider the formation and settling of dust particles in a self-consistent way, e.g. the spatial extent, microphysics, and composition of the dust cloud is determined by the fundamental atmospheric parameters ($T_\mathrm{eff}$, $\log\,g$, composition). The particle sizes change with the cloud depth and the atmospheric parameters, ranging from sub-micron to $10^{2}\mu$m sizes. The chemical model accounts for the non-equilibrium chemistry of CO, CH$_{4}$, N$_{2}$, NH$_{3}$, and CO$_{2}$.  
\end{itemize}

\begin{table*}[t]
\label{Tab:demod}
\centering
\caption{\label{Tab:atmomodchar} Characteristics of the atmospheric model grids fitted on the planet full SED.}
\begin{tabular}{llllllll}
\hline
Model name & $T_{\mathrm{eff}}$& $\Delta T_{\mathrm{eff}}$ &  log $g$ & $\Delta$ log $g$ & $[M/H]$ & $[\alpha]$  \\
	& (K) & (K) & (dex)  & (dex)  & (dex)  & (dex)  \\
\hline \hline
BT-SETTL14 & 500-3000 & 50 & 3.5-5.5 & 0.5 & 0.0 & 0.0 \\
 & 500-2800 & 100 & 4.0-5.5 & 0.5 & 0.3 & 0.0 \\ 
\hline 
Exo-REM  & 700-2000 & 100 & 2.1-5.3 & 0.1 &  -0.5,0,0.5 & 0.0 \\
\hline
Cloud AE-60		& 600-1025 & 25,50 & 3.5-5.0	& 0.1,0.3,0.5	&	0.0	& 0.0 \\
		& 1050-1700 & 100 & 3.5-5.0	& 0.5	&	0.0	& 0.0 \\
\hline
\end{tabular}
\end{table*}

The  coverage of each model grid is reported in Table~\ref{Tab:atmomodchar}. The Exo-REM4 and BT-SETTL grids  enable to explore the effect of metallicity. We used the models with solar metallicity to compare the results obtained with the different grids first, and then explored the effect of metallicity later on. \\ 

We considered a joint fit of the available spectra, and non-overlapping broad and narrow-band photometry of the four planets. The  fluxes obtained from the synthetic spectra (fluxes emitted per surface unit at the top of the atmosphere) were fitted onto the apparent fluxes of the planets using a dilution factor which scales with  the radius of the objects.  

In order to account for the non-homogeneous wavelength sampling of the photometry and spectra and for the filter widths, we used the G goodness-of-fit indicator proposed by \cite{2008ApJ...678.1372C}. Our implementation of the method is described in Vigan et al. 2015, submitted.  We considered a correlated error on the absolute photometry of the P1640 spectrum of HR8799c  and OSIRIS spectrum of HR8799b of 0.13 and 0.11 mag, respectively. This corresponds to the uncertainty on the broad-band photometry of the planets used to flux-calibrate their spectra.  We derived the three most probable fitting solutions inferred from the Monte-Carlo simulations \citep[$f_{MC}$ indicator, see][]{2008ApJ...678.1372C} and kept the one producing the best visual fit. 

We also performed a comparison considering a standard $\chi^{2}$. In that case, the fit is mostly sensitive to the spectral shape, and not so much on the photometry which complements the SEDs of the planets. This means that the fit is sensitive to a wavelength range where the effect of dust opacity dominates, and not so much to non-equilibrium chemistry effects which dominate longward of 2.5 $\mu$m.

\subsection{Results}

\begin{figure*}
\begin{center}
\includegraphics[width=13cm]{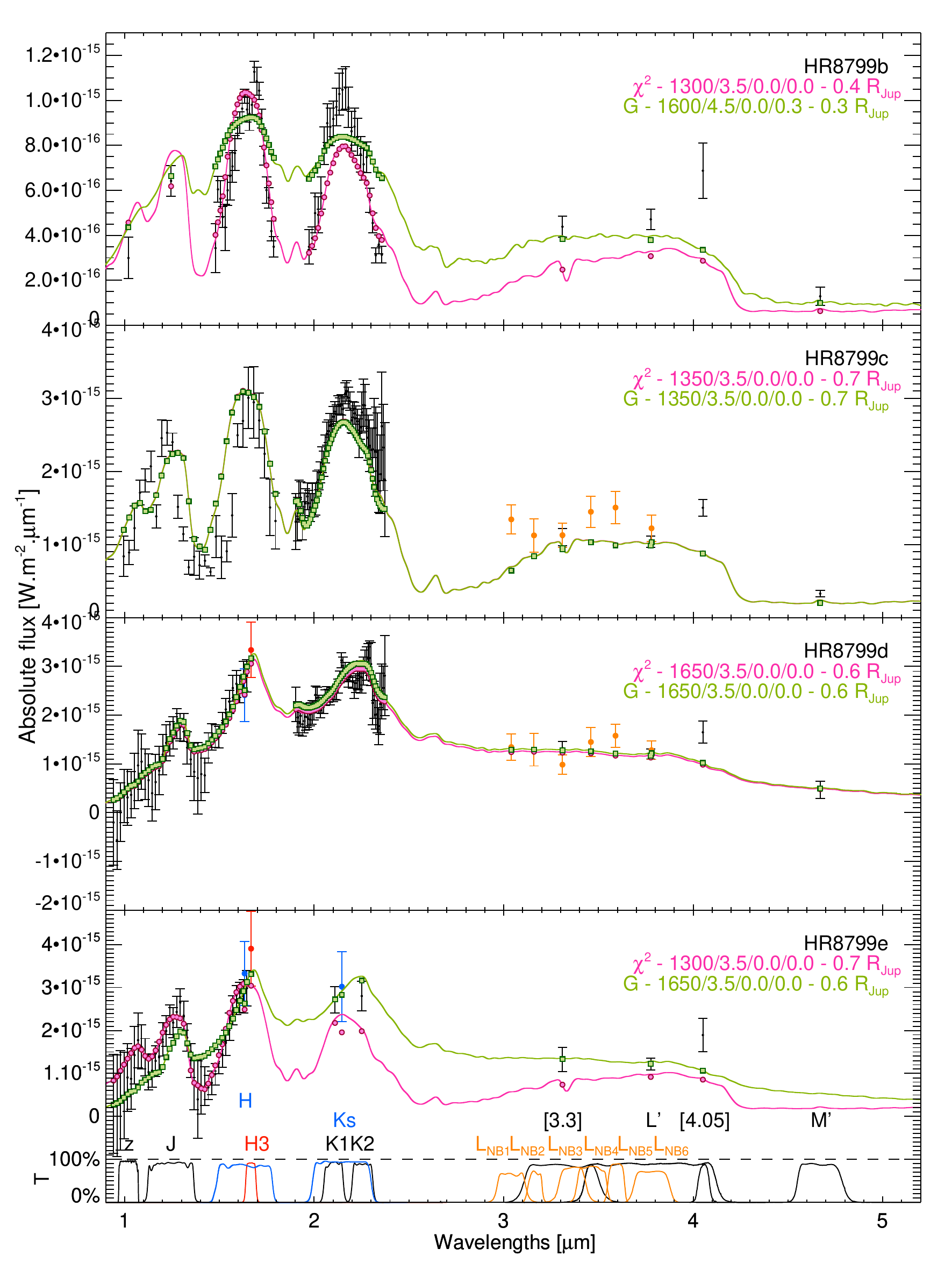}
\caption{Comparison of the SEDs of HR8799bcde to the best-fitting BT-SETTL14 synthetic spectra found minimizing the G and $\chi^{2}$ goodness-of-fit indicators. The corresponding atmospheric parameters $T_{eff}/log\:g/[M/H]/[\alpha]$ are reported in each panel.}
\label{Fig:SEDsynthSETTL}
\end{center}
\end{figure*}

\begin{figure*}
\begin{center}
\includegraphics[width=13cm]{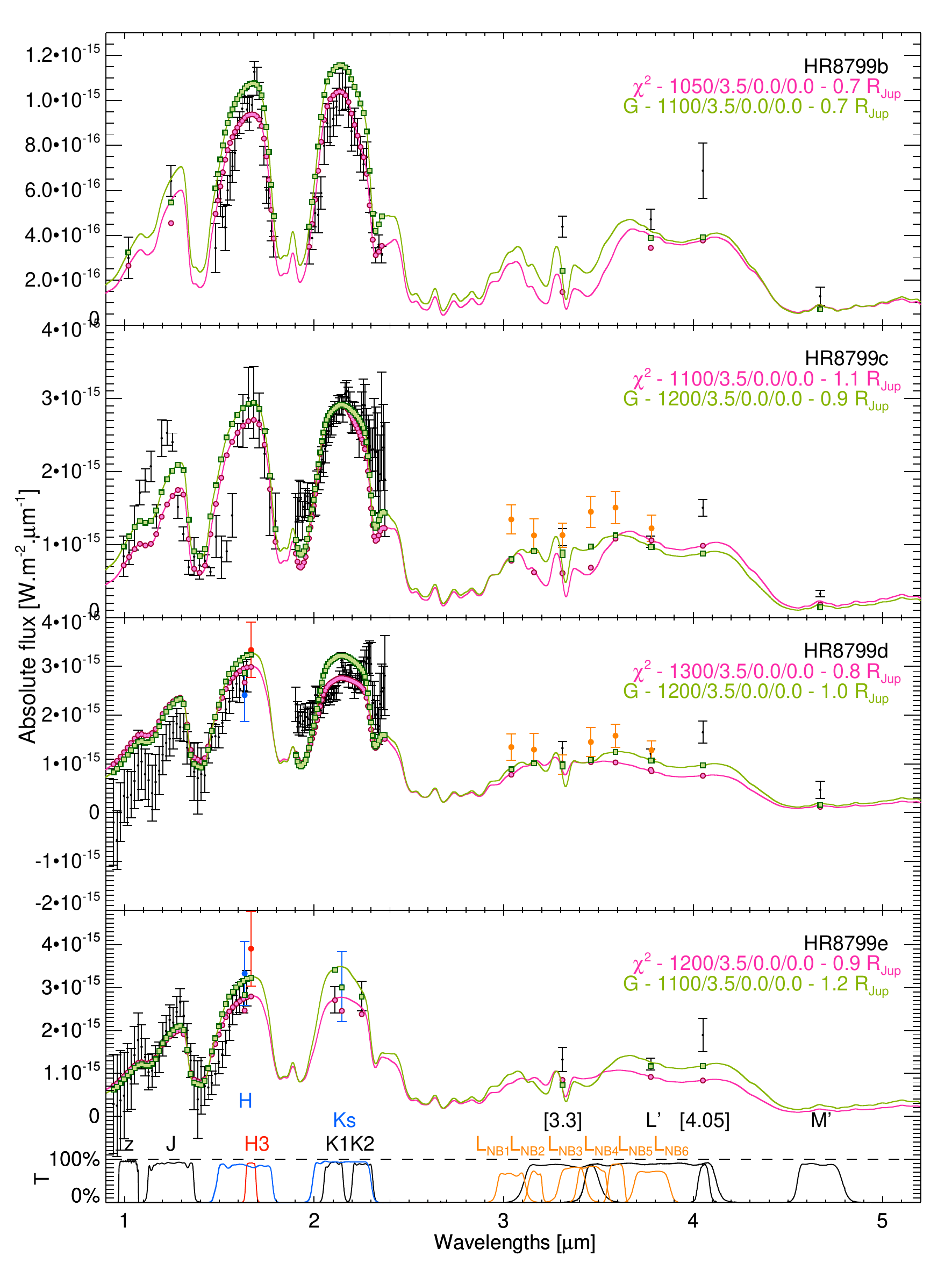}
\caption{Same as Figure \ref{Fig:SEDsynthSETTL} but for the Cloud AE-60 models from \cite{2011ApJ...737...34M}.}
\label{Fig:SEDsynthBurrows}
\end{center}
\end{figure*}

\begin{figure*}
\begin{center}
\includegraphics[width=13cm]{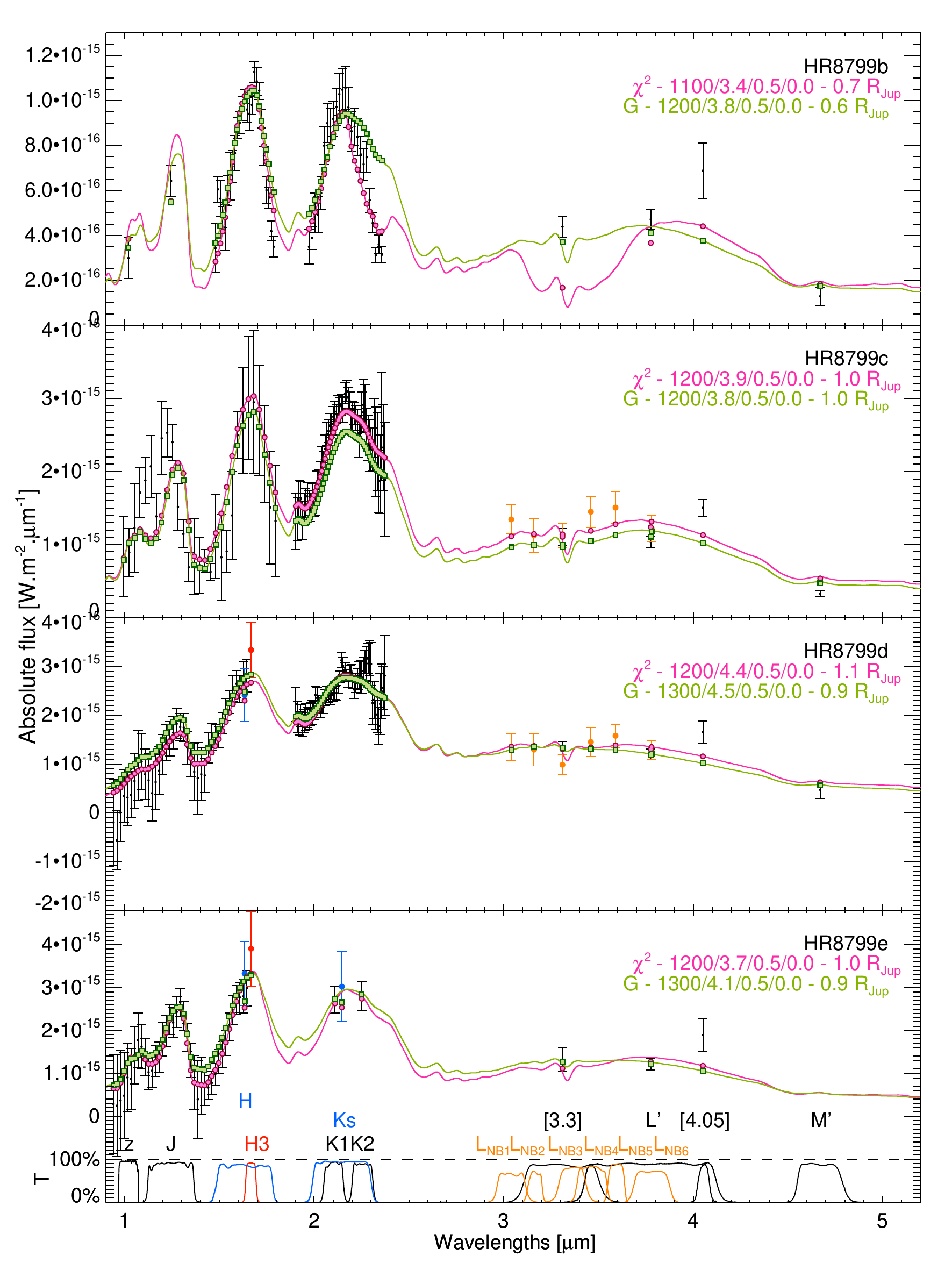}
\caption{Same as Figure \ref{Fig:SEDsynthSETTL} but for the Exo-REM models.}
\label{Fig:SEDsynthEXOREM}
\end{center}
\end{figure*}

The best-fitting parameters, including the dilution radii are reported in Table \ref{Tab:SEDsynthpar}.  We only give the best solution found  among the five grids of Exo-REM model. We plot in Figures \ref{Fig:SEDsynthSETTL}, \ref{Fig:SEDsynthBurrows}, and \ref{Fig:SEDsynthEXOREM} the best fitting solutions for the BT-SETTL14, Exo-REM, and Cloud AE-60 models, respectively. 

\begin{table*}[t]
\centering
\caption{\label{Tab:SEDsynthpar} Fitting solutions for HR8799bcde spectral-energy distributions and the three sets of atmospheric models}
\begin{tabular}{lllllllll}
 \hline
 Planet & Model name   & $T\mathrm{_{eff}}$    &  log $g$  & [M/H]   & R & G & $\chi_{red}^{2}$ & $f_{MC}$ \\
              &  			      & (K)   & (dex)  & (dex)  & ($R_{Jup}$) & & &  \\
 \hline
HR8799 b - G &  BT-SETTL14   &  1600 & 4.5   & 0.3  & 0.3 & 5.78 & n.a. &  0.55\\
					 &  Exo-REM4 & 1200 & 3.8	& 0.5  & 0.6 & 4.03 & n.a. & 0.15 \\ 
					 &  Cloud AE-60	& 1100 & 3.5 & 0.0  & 0.7 & 6.78 & n.a. & 0.13 \\ 
HR8799 b - $\chi^{2}$			    &  BT-SETTL14   & 1300  & 3.5  & 0.0 & 0.4 & n.a. & 4.43 & n.a.   \\ 
											&  Exo-REM4 & 1100  & 3.4  & 0.5  & 0.7 & n.a. & 3.76 & n.a.   \\ 
											&  Cloud AE-60	& 1050 & 3.5  & 0.0 & 0.7 & n.a. & 3.82 & n.a.   \\ 
 \hline 
HR8799 c - G &  BT-SETTL14   &  1350 & 3.5 & 0.0  & 0.7 & 6.74 & n.a. &  0.18\\ 
					&  Exo-REM4 & 1200 & 3.8	& 0.5   & 1.0 & 4.20 & n.a. & 0.33 \\
					&  Cloud AE-60	& 1200 & 3.5 & 0.0 & 0.9 & 7.63 & n.a. & 0.75 \\
HR8799 c - $\chi^{2}$			    &  BT-SETTL14   & 1350  & 3.5   & 0.0 & 0.7 & n.a. & 7.80 & n.a.   \\ 
											&  Exo-REM4 & 1200  & 3.9  & 0.5 & 1.0 & n.a. & 4.33 & n.a.   \\ 
											&  Cloud AE-60	& 1100 & 3.5 & 0.0  & 1.1 & n.a. & 7.51 & n.a.   \\ 
 \hline 
HR8799 d - G &  BT-SETTL14   &  1650 & 3.5  & 0.0   & 0.6 & 0.49 & n.a. &  0.78\\ 
					 &  Exo-REM4 & 1300 & 4.5	& 0.5  & 0.9 & 0.51 & n.a. & 0.13 \\
				 	 &  Cloud AE-60	& 1200 & 3.5 & 0.0  & 1.0 & 3.20 & n.a. & 0.85 \\
HR8799 d - $\chi^{2}$			    &  BT-SETTL14   & 1650  & 3.5  & 0.0  & 0.6 & n.a. & 0.81 & n.a.   \\  
				&  Exo-REM4 & 1200 & 4.4	& 0.5  & 1.1 & n.a. & 0.84 & n.a. \\
				& Cloud AE-60	& 1300 & 3.5 & 0.0  & 0.8 & n.a. & 3.88 & n.a. \\
 \hline
HR8799 e - G &  BT-SETTL14   &  1650 & 3.5  & 0.0   & 0.6 & 0.63 & n.a. &  0.56\\ 
			    &  Exo-REM4 & 1300 & 4.1	& 0.5  & 0.9 & 0.32 & n.a. & 0.13 \\
			    &  Cloud AE-60	& 1100 & 3.5 & 0.0 & 1.2 & 1.42 & n.a. & 0.29 \\
HR8799 e - $\chi^{2}$			    &  BT-SETTL14   & 1300  & 3.5  & 0.0  & 0.7 & n.a. & 0.94 & n.a.   \\ 
				&  Exo-REM4 & 1200 & 3.7	& 0.5 & 1.0 & n.a. & 0.28 & n.a. \\
				& Cloud AE-60	& 1200 & 3.5 & 0.0  & 0.9 & n.a. & 1.50 & n.a. \\
 \hline
\end{tabular}
\end{table*}

Our analysis reveals that the BT-SETTL models fail to reproduce  the shape and the absolute fluxes of the SED of each planets simultaneously. The fitted temperatures are too high compared to what is expected from the luminosity of the object, or equivalently they produce fits with unphysical sub-Jupiter radii. This is consistent with the conclusions  derived for the dusty L7 dwarfs WISEP J004701.06+680352.1 and PSO J318.5338-22.8603 \citep{2013ApJ...777L..20L, 2015ApJ...799..203G}, e.g. the two objects whose SED represent the best the ones of HR8799d and e (Section \ref{subsec:empirical}). In addition, the models do not provide a good fit of the datapoints in the 3-4 $\mu$m regime while they are the only ones that account for non-equilibrium chemistry. \cite{2014A&A...564A..55M} suggested that  the BT-SETTL models do not seem to produce naturally enough dust, especially at high altitude in the cloud. The problem could be emphasized here with the HR8799 planets. 

In that respect, the fit of HR8799bcde is much improved using the two sets of "thick cloud" models. This is in agreement with the conclusions from previous studies \citep[][and ref. therein]{2011ApJ...737...34M, 2012ApJ...754..135M, 2014ApJ...795..133C, 2014ApJ...792...17S}. However, we can still notice that the Cloud AE-60 models do not have a pseudo-continuum slope red enough to reproduce simultaneously the whole SEDs. Their water-band absorption are too deep compared to the ones of the planets. In comparison, \cite{2011ApJ...737...34M} find $T\mathrm{_{eff}}$ 100 to 350K lower than the ones we estimate, based on a $\chi^{2}$ fitting process and the same models. These differences may arise from the lower number of photometric points used to perform the fit and to the fitting procedures. 

The remaining issues found with the Cloud AE-60 models are solved by the Exo-REM models. These models fit within 2$\sigma$ the whole set of photometric datapoints of the HR8799d and e simultaneously at solar metallicity. The best fits are always obtained with the  Exo-REM4 grid, which corresponds to models with the thicker clouds. The whole fit is improved -- in particular the water band absorptions -- for the four planets using models with super-solar metallicity (M/H=0.5 dex). Our $\chi^{2}$ maps indicate that the constraint onto this parameter is not significant ($<1\sigma$). But  88\%, 79\%, 100\%, and 77\% of the solutions found for HR8799e, d, c, and b with the G estimator are for M/H=+0.5.  The models successfully reproduce the spectrophotometric properties of HR8799d and e for $T_{\mathrm{eff}}=1200$ K and log $g$ typical of young  objects \citep[e.g. ][]{2014A&A...562A.127B} and compatible with previous studies of the planets \citep[][and ref therein]{2012ApJ...754..135M}. The best fitting log $g$ increases when models with increasing M/H are used, as expected \citep{2007ApJ...657.1064M}.

The Exo-REM models provide radii around 1 $R_{Jup}$ which are still 20 to 30$\%$ smaller than those expected from evolutionary models \citep{2003A&A...402..701B}, especially for HR8799b. We attempted a fit of the SEDs with a restrained grid  (800 $\leq T_{\mathrm{eff}} \leq$ 1500, 3.0 $\leq$ log $g$ $\leq$ 4.5)  of Exo-REM models with extremely thick clouds ($\tau_{ref}=5$), a mean particle size of 30 $\mu$m, and solar metallicity (hereafter Exo-REM 6). The fit is slightly improved (lower $\chi^{2}$ and G values) for HR8799b and d, but not for the c and e planets. Moreover, the determined $T\mathrm{_{eff}}$ and radii remain unchanged.  But we find that the determined log $g$ decreases when the content of dust increases (models 1 to 6, exept for model 5). It is consistent with the trend seen when increasing the metallicity for the Exo-REM4 models. This exploration of the model degeneracies calls for more finetuning of the cloud properties \citep[e.g. similar to ][]{2012ApJ...754..135M, 2013ApJ...778...97L}, considering cases with different dust size distributions and fully exploring the impact of metallicity.  We also note that despite the reasonable fit of the SEDs found here,  the Exo-REM models should be validated  on field L and T dwarf objects.

In summary, we find that the Exo-REM atmospheric models with thick clouds provide a simultaneous fit  (within 2$\sigma$) of the SED  of HR8799de for  $T_{eff}=1200$ K and log $g$ (3-4.5 dex). These temperatures are $\sim$100-200 K higher than the ones found so far \citep[see the summary in ][]{2014ApJ...792...17S} and to evolutionary models predictions (from the luminosity estimate). This and the worse fit for HR8799b imply that the cloud modelling still needs to be improved.

\section{Discussion}
\label{sec:discu}
\subsection{Are the HR8799 planets properties so peculiar?}
\label{subsec:var}
The conclusions of the empirical analysis are in line with recent results reporting good matches between known directly imaged young companions and low gravity field dwarf spectra  \citep{2013ApJ...772...79A}, of which some are candidate members of young moving groups \citep{2014ApJ...783..121G}. For instance, the presently available spectra of the exoplanet $\beta$ Pictoris b are known to be reproduced by those of young L0-L1 dwarfs \citep{2014A&A...567L...9B, 2015ApJ...798L...3C}.  The same happens for AB Pic b \citep{2010A&A...512A..52B} or 1RXS J160929.1-210524 b \citep{2014A&A...564A..55M}. 

Previous studies of the HR8799 planets have also proposed that the photometric and spectroscopic properties of the objects could be explained by 1/ disequilibrium chemistry \citep{2011ApJ...733...65B,2012ApJ...754..135M,2012ApJ...753...14S, 2014ApJ...792...17S,2014ApJ...794L..15I}, and/or 2/ patchy atmospheres \citep{2012ApJ...754..135M, 2012ApJ...753...14S, 2014ApJ...792...17S}, and/or 3/ non-solar compositions \citep{2011ApJ...733...65B, 2013Sci...339.1398K, 2013ApJ...778...97L, 2015ApJ...804...61B}.  The good fit in the 3-5 $\mu$m region for the planets d and e (along with the other wavelengths), despite the lack of non-equilibrium chemistry, non-solar metallicity compositions, and local variation of the cloud thickness in the models, suggest that these ingredients are not critically needed here.  Our analysis rather demonstrates that a higher content in dust and reduced CIA of $H_{2}$ resulting from a low surface gravity can explain the deviation of the planet's SED with respect to those of standard field L and T dwarfs.  

 This does however not mean that such additional ingredients do not play a role, especially for HR8799b and c.  \citet{2011ApJ...733...65B} noted that the $K$ band spectrum of the high-amplitude variable brown dwarf T1 dwarf \object{2MASS J21392676+0220226} fitted well the SED of HR\,8799\,b. The spectrophotometric properties of  the two objects representing the best the properties of HR\,8799\,b and c, \object{2MASS J13243553+6358281} and \object{SDSS J151643.01+305344.4}, can be modeled by blended spectra of L-T dwarfs, e.g. as unresolved L/T transition binary at field dwarf ages \citep{2010ApJ...710.1142B, 2011ApJ...732...56G}. Beside its possible young age, we can speculate that \object{SDSS J151643.01+305344.4} is a good template because the integrated flux of the possible binaries mimics layered patchy atmospheres. Conversely, \cite{2014arXiv1412.6733H} find that \object{2MASS J13243553+6358281} experiences high amplitude variability in the red-optical. 

The planets e and d could also be variable objets. The best fitting template to HR8799d's SED -- \object{PSO J318.\-53\-38-22.86\-03}  -- was recently found to experience photometric variability in the near-infrared (Biller et al. 2015, submitted to ApJ Letter). In addition, the red late-L dwarf WISE1738, which reproduces well the properties of HR8799d and e also has a highly variable near-infrared spectrum \citep{2013ApJS..205....6M}. 

In that sense, the good fit of the planet spectra with field dwarfs templates reddened by sub-micron iron, corundum, or silicate grain opacities  in Section~\ref{subsec:empirical} is  in line with the conclusions of \cite{2015ApJ...798L..13Y} who find that the observed variability of L5 dwarfs can be explained by the presence of spatially varying high-altitude haze layers above the condensate clouds. Such a layer may exist into the planet atmospheres and could be added to the current self-consistent cloud models (e.g. BT-SETTL, DRIFT-PHOENIX) to bring the $T\mathrm{_{eff}}$ and radii estimate in agreement with the expectations from evolutionary models.

\subsection{Masses and diversity among the HR8799 system}
\cite{2008Sci...322.1348M}, \cite{2010Natur.468.1080M}, and \cite{2011ApJ...736L..33C} used the bolometric luminosity of the planets to estimate masses of 5,7,7, and 7 $M_{Jup}$ for HR8799b,c,d and e, respectively if the system is 30 Myr old. We retrieve the same luminosity -- and then mass estimate -- for HR8799d and e using the bolometric correction estimated for the dusty L7 dwarf WISE0047 by \cite{2015ApJ...799..203G}.  The radii and log $g$ derived from the best atmospheric model fit yield masses marginally consistent with these values. This is most likely due to the incomplete exploration of the degeneracies between log $g$, $T\mathrm{_{eff}}$,  M/H, and the cloud properties, and also to remaining uncertainties in the models. A revision of these mass estimate based on evolutionary models coupling the interior with atmospheres containing thick cloud layers would also be needed \citep[e.g.][]{2011ApJ...736...47B, 2012ApJ...754..135M}. 

It is worth comparing the original mass estimates of the planets to the conclusions from the empirical analysis. The planets d and e have very close SEDs and are both fitted by dusty L7 dwarfs. The ages of the two best fitting templates, and the position of the objects in the color-color diagrams with respect to the reddening vectors produced by silicate, corundum, and iron grains, suggest that the remaining differences between the SED of the two planets may come from slight differences on the cloud properties and atmospheric parameters (as suggested by the Exo-REM models).  On the contrary, the deeper water-band absorption of HR8799c and the semi-empirical fit rather indicate that this planet  has a  later spectral type. HR8799b pursues this trend (Section \ref{subsec:empirical}). The COND hot-start evolutionary models \citep{2003A&A...402..701B} predict that the log $g$ of the four planets should be nearly the same (3.90-4.04 dex) assuming that the mass can be safely derived from the object luminosity and cooling tracks. Therefore, this suggests that 1/ a tiny change in log $g$ and/or $T\mathrm{_{eff}}$ and/or composition in-between the planets d and e and the planets  b and c is also responsible for the observed differences, or 2/ the $T\mathrm{_{eff}}$ -- therefore the mass if the hot-start scenario holds -- of planets d and e are identical, but the $T\mathrm{_{eff}}$  and mass of HR8799c and then b are lower. 

\section{Conclusions}
We used a compilation of SPHERE photometry and spectra in addition to 1-5 $\mu$m photometry of the four planets around HR8799 to reinvestigate the properties of these objects. We demonstrate that the  1-5 $\mu$m spectro-photometric properties of HR\,8799\,d and e are similar to those of  the population of dusty L dwarfs with estimated spectral type L6-L8 dwarfs that start to be unearthed in the solar neighborough. We show in addition that the spectrophotometric properties of HR8799b and c are reproduced exclusively by the SED of proposed young, and/or binary candidates, and/or variable early-T brown dwarfs that we reddened by the opacity of sub-micronic  silicate (enstatite, forsterite), corundum, or iron grains, e.g. species expected to be abundant in the atmosphere of brown dwarf and young warm giant planet atmospheres.

We use the BT-SETTL, Exo-REM, and \cite{2011ApJ...737...34M} models to confirm that  models with thick clouds  reproduce better the  1-5 $\mu$m flux and spectral shape of the planets. Despite the increased number of photometric points, the Exo-REM models fit the whole SED of HR8799d and e within 2$\sigma$. They provide $T\mathrm{_{eff}}$ in the range 1100-1300 K for HR8799d and e and fitted radii almost consistent with predictions from hot-start evolutionary models.  The models still fail to reproduce the shape and absolute fluxes of HR8799 b and c. Because the Exo-REM models do not consider non-equilibrium chemistry and patchy atmospheres, we conclude that these two ingredients are not critically needed to explain the properties of HR8799d and e.  Instead, the analysis confirms that the atmosphere of these planets is heavily affected by dust opacity.
 
The long-slit spectroscopic mode of SPHERE should soon be able to provide higher resolution spectra (R$\sim$400) of the planets from 1 to 1.8 $\mu$m and extend the spectra coverage to the K-band, as shown by \cite{2015ApJ...805L..10H} and Maire et al.  (submitted to A\&A). The ALES integral field spectrograph  (currently being installed on the LBT/LMIRCam instrument) will also be able to extend the  spectral coverage of the planets up to 5 $\mu$m \citep{2015arXiv150806290S}. Together, the new data could bring additional constraints on the surface gravity, cloud properties, enrichments at formation, and role of non-equibrium chemistry in the atmosphere of these planets.
\label{sec:conc}

\begin{acknowledgements}
We are grateful to the SPHERE team and all the people at Paranal for the great effort during SPHERE commissioning runs. We thank Ben Oppenheimer, Laurent Pueyo, and Travis Barman for providing an access to their spectra of the HR8799 planets. We also thank P. Ingraham, G. Mace, J. Gizis, M. Liu, A. Schneider, D. Stephens, M. Cushing, B. Bowler, D. Kirkpatrick,  P. Rojo, K. Allers, J. Patience, M.-E. Naud, and B. Gauza for sending us their spectra of young companions and dusty dwarfs. This research has benefitted from the SpeX Prism Spectral Libraries, maintained by Adam Burgasser at http://pono.ucsd.edu/$\sim$adam/browndwarfs/spexprism. M.B., G. Ch, A.-M. L., J.-L. B., F. A., D. H. and D. M.  acknowledge support from the  French National Research Agency (ANR) through the GUEPARD project grant ANR10-BLANC0504-01 and from the Programmes Nationaux de Planétologie et de Physique Stellaire (PNP \& PNPS, CNRS/INSU). A.Z., D.M., R.G., R.C., and S.D. acknowledge partial support from PRIN INAF 2010 ``Planetary systems at young ages''. A.Z., D.M., A.-L.M., R.G., S.D., and R.U.C.  acknowledge support from the ``Progetti Premiali'' funding scheme of the Italian Ministry of Education, University, and Research.  A. Z. acknowledges support from the MillenniumvScience Initiative (Chilean Ministry of Economy), through grant ``Nucleus RC130007''.  SPHERE is an instrument designed and built by a consortium consisting of IPAG (Grenoble, France), MPIA (Heidelberg, Germany), LAM (Marseille, France), LESIA (Paris, France), Laboratoire Lagrange (Nice, France), INAF – Osservatorio di Padova (Italy), Observatoire de Gee\`{e}ve (Switzerland), ETH Zurich (Switzerland), NOVA (Netherlands), ONERA (France) and ASTRON (Netherlands), in collaboration with ESO. SPHERE was funded by ESO, with additional contributions from CNRS (France), MPIA (Germany), INAF (Italy), FINES (Switzerland) and NOVA (Netherlands).  SPHERE also received funding from the European Commission Sixth and Seventh Framework Programmes as part of the Optical Infrared Coordination Network for Astronomy (OPTICON) under grant number RII3-Ct-2004-001566 for FP6 (2004–2008), grant number 226604 for FP7 (2009–2012) and grant number 312430 for FP7 (2013–2016). DH acknowledges support from the European Research Council under the European Community’s Seventh Framework Programme (FP7/2007-2013 Grant Agreement No. 247060) and from the Collaborative Research Centre SFB 881 ``The Milky Way System'' (subproject A4)
of the German Research Foundation (DFG). 
\end{acknowledgements}

\bibliographystyle{aa}
\bibliography{hr8799}

\end{document}